\documentclass[journal,twoside,web]{ieeecolor}
\usepackage{cite}
\usepackage{color}
\usepackage{amsmath,amssymb,amsfonts}
\usepackage{algorithmic}
\usepackage{graphicx}
\usepackage{textcomp}
\usepackage{multicol}
\usepackage{booktabs}
\usepackage{multirow}
\usepackage{colortbl}
\def\BibTeX{{\rm B\kern-.05em{\sc i\kern-.025em b}\kern-.08em
    T\kern-.1667em\lower.7ex\hbox{E}\kern-.125emX}}
\begin{document}
\title{Annotation-Efficient Learning for Medical Image Segmentation based on Noisy Pseudo Labels and Adversarial Learning}
\author{Lu Wang, Dong Guo, Guotai Wang and Shaoting Zhang
\thanks{This work was supported  in part by the National Natural Science Foundation of China [81771921, 61901084], in part by the Key Research and Development Project of Sichuan  Province [20ZDYF2817] and in part by the Science and Technology Commission of Shanghai Municipality [19511121400]. (Corresponding author: Guotai Wang)}
\thanks{L. Wang, D. Guo, G. Wang and S. Zhang are with the School of Mechanical and Electrical
Engineering, University of Electronic Science and Technology of China,
Chengdu, 611731, China. S. Zhang is also with SenseTime Research, Shanghai, 200233, China. (email: guotai.wang@uestc.edu.cn)}
\thanks{Copyright of this paper has been transferred to IEEE. The published version is available at: https://doi.org/10.1109/TMI.2020.3047807.}
}
\maketitle

\begin{abstract}
Despite that deep learning has achieved state-of-the-art performance for medical image segmentation, its success relies on a large set of manually annotated images for training that are expensive to acquire. In this paper, we propose an annotation-efficient learning framework for segmentation tasks that avoids annotations of training images, where we use an improved Cycle-Consistent Generative Adversarial Network (GAN) to learn from a set of unpaired medical images and auxiliary masks obtained either from a shape model or public datasets. We first use the GAN to generate pseudo labels for our training images under the implicit high-level shape constraint represented by a Variational Auto-encoder (VAE)-based discriminator with the help of the auxiliary masks, and build a Discriminator-guided Generator Channel Calibration (DGCC) module which employs our discriminator's feedback to calibrate the generator for better pseudo labels. To learn from the pseudo labels that are noisy, we further introduce a noise-robust iterative  learning method using noise-weighted Dice loss. We validated our framework with two situations: objects with a simple shape model like optic disc in fundus images and fetal head in ultrasound images, and complex structures like lung in X-Ray images and liver in CT images. Experimental results demonstrated that 1) Our VAE-based discriminator and DGCC module help to obtain high-quality pseudo labels. 2) Our proposed noise-robust learning method can effectively overcome the effect of noisy pseudo labels. 3) The segmentation performance of our method without using annotations of training images is close or even comparable to that of learning from human annotations. 
\end{abstract}


\section{Introduction}
\label{sec:introduction}
\IEEEPARstart{M}{edical} image segmentation is important for a wide range of clinical applications~\cite{shen2017deep}, such as modeling of organs, accurate diagnosis, quantitative measurement and surgical planning for tumors. Nowadays, deep learning with Convolutional Neural Networks (CNNs) has achieved great success for medical image segmentation tasks~\cite{shen2017deep}, such as segmentation of fetal head~\cite{wu2017cascaded}, optic disc~\cite{sevastopolsky2017optic}, brain tumor~\cite{wang2017automatic} and pancreas~\cite{roth2015deeporgan}. Their success highly depends on the availability of a large set of training images with manual annotations given by experts. However, a large set of manual annotations for medical image segmentation are difficult to acquire as giving pixel-level annotations for segmentation tasks is time-consuming and relies on experts with domain knowledge to implement. Therefore, it is expensive and labor-intensive to acquire high-quality manual annotations for training, which has become the main obstacle for developing deep leaning models for medical image segmentation tasks~\cite{weese2016four}.

To tackle this issue, annotation-efficient learning for medical image segmentation has attracted increasing attention as it helps to reduce the requirement of large amount of annotations for training~\cite{tajbakhsh2020embracing}. First, some weakly-supervised methods have been proposed where the deep learning model is only trained with image-level labels~\cite{feng2017discriminative}, sparse pixel-level annotations (e.g., scribbles) and bounding boxes~\cite{rajchl2016deepcut}. Second, to avoid annotating the entire dataset, semi-supervised methods~\cite{bai2017semi,nie2018asdnet} have been proposed for segmentation where only a subset of the images are annotated. In addition, some intelligent interactive segmentation/annotation tools~\cite{8370732,8270673} have also been developed to reduce the human efforts for pixel-level annotation. Despite their values in alleviating challenges of acquiring large-scale human annotations, these methods still require a lot of efforts of human annotators and are not free from human annotations of training images.   

Avoiding annotations of training images has a potential to further overcome the difficulty and high cost of acquiring a large annotated training set. As an attempt towards this goal, unsupervised cross-modality adaptation methods~\cite{jiang2018tumor,chen2020unsupervised} are proposed to remove the need of annotations in one modality (i.e., target domain) given annotated images in another modality (i.e., 
source domain), but they still require annotations in the source domain. Some traditional unsupervised methods that do not require human annotations for training such as the Iterative Randomized Hough Transform (IRHT)~\cite{lu2008detection} and texture-based ellipse detection~\cite{perez2015automatic} were proposed to detect ellipse-like fetal head from ultrasound images, and they have a low robustness when dealing with images with weak boundary information. Currently, there have only been few works on learning a segmentation model without the use of annotations of training images. For example, in~\cite{moriya2018unsupervised,moriya2019unsupervised}, deep representation learning is proposed for unsupervised 3D medical image segmentation. However,  the performance of such a method is still much lower than learning from human annotations. 

In this work, we propose a novel annotation-efficient deep learning framework for medical image segmentation. Our core idea is to learn from a set of auxiliary object masks that are unpaired with training images and can be easily obtained through either shape prior information or publicly available datasets in a probably different domain. For some well-shaped objects that can be accurately described by a parametric model, we directly use the parametric model as shape prior to generate a set of auxiliary masks. For objects with more complex shapes that can hardly be fitted by a parametric model, we take advantage of object mask samples from other available domains (e.g., public datasets). Based on the unpaired set of training images and auxiliary masks, we use a cycle-consistent Generative Adversarial Network (CycleGAN) where a generator learns to obtain pseudo labels of training images. Unlike the work in~\cite{moriya2018unsupervised,moriya2019unsupervised} that assigns labels using deep feature representation and clustering for unsupervised segmentation, our method introduces implicit shape constraints through adversarial learning with auxiliary masks to obtain more accurate pseudo labels.
Based on the noisy pseudo labels, we propose a noise-robust iterative training procedure with a noise-weighted Dice loss to train a final segmentation model to achieve high segmentation performance. Therefore, the entire training process does not require manual annotations for corresponding images in our training set.

\subsection{Contributions}
The contributions of this work are three-fold. First, we propose a novel annotation-efficient deep learning framework for medical image segmentation, where the model learns from a set of auxiliary masks that can be easily obtained and unpaired with our training images, thus the manual annotation for each training image is not required. The framework consists of a novel pseudo label generation module that obtains an initial pseudo segmentation label for each training image and an iterative learning module that is robust against noise in the initial pseudo labels. Second, to obtain high-quality pseudo labels, we propose a VAE-based discriminator that encourages a high-level shape constraint on the pseudo labels and propose a Discriminator-guided Generator Channel Calibration (DGCC) module to calibrate the channel-wise information of pseudo label generator using the discriminator's feedback. Thirdly, we propose a novel iterative noise-robust training method to learn from the pseudo labels, where low-quality pseudo labels are rejected by a Label Quality-based Sample Selection (LQSS) module and a noise-weighted Dice loss is proposed to boost the performance of the final segmentation model. Experiments showed that for optic disc segmentation and fetal head segmentation, our method achieved close or even comparable performance to learning from human annotations. For more complex objects such as the lung and the liver, our annotation-efficient method can also achieve very competitive results and outperform existing unsupervised, weekly supervised and domain adaptation-based methods.

\subsection{Related Works}
\subsubsection{Deep Learning for Medical Image Segmentation}
Recent advances in medical image segmentation are based on CNNs~\cite{shen2017deep}, such as U-Net~\cite{ronneberger2015u,cciccek20163d} and V-Net~\cite{milletari2016v}. Some more powerful networks including Attention U-Net~\cite{oktay2018attention}, U-Net++~\cite{zhou2018unet++} and H-DenseUNet~\cite{li2018h} further improved the segmentation performance for several tasks. However, most existing works require a large set of manual annotations for training. 
\subsubsection{Annotation-Efficient Learning}
Existing annotation-efficient methods for medical image segmentation mainly include weakly- and semi-supervised learning, unsupervised domain adaptation and unsupervised learning~\cite{tajbakhsh2020embracing}.

For weakly-supervised deep learning, Feng et al.~\cite{feng2017discriminative} used image-level labels to train a pulmonary nodule classification network, and used its activation map for segmentation of lung nodules. Rajchl el al.~\cite{rajchl2016deepcut} combined GrabCut~\cite{rother2004grabcut} and iterative training for fetal brain segmentation using bounding box annotations.
Semi-supervised learning methods only require a subset of training images to be labeled~\cite{xie2019semi}. Bai et al.~\cite{bai2017semi} proposed to alternatively update the network parameters and the segmentation for the unlabeled data for cardiac MR image segmentation. Nie et al.~\cite{nie2018asdnet} used a region attention and adversarial learning-based method for this purpose. Other methods such as self-ensembling~\cite{yu2019uncertainty} and consistency loss~\cite{mittal2019semi} have also been proposed for semi-supervised segmentation.
Unsupervised domain adaptation is practically appealing where no label is available for the target domain with available annotations for the source domain. With the great success of CycleGAN~\cite{zhu2017unpaired} in unpaired image-to-image translation, many approaches~\cite{jiang2018tumor,chen2020unsupervised} used CycleGAN to transform target domain images to the source domain for the segmentation. Jiang et al.~\cite{jiang2018tumor} used CycleGAN with tumor-aware loss to transform CT images to MRI images for lung cancer segmentation where annotations for MRI images were not available. Chen et al.~\cite{chen2020unsupervised} synergistically exploited feature alignments with image transformation to deal with the domain adaptation between CT and MRI for cardiac substructure and abdominal organ segmentation.

For training segmentation models without any human annotations, Moriya et al.~\cite{moriya2018unsupervised} used deep feature representations of training patches and clustering for unsupervised segmentation. Moriya et al.~\cite{moriya2019unsupervised} employed adversarial learning with categorical latent variables for unsupervised segmentation of micro-CT images. However, these methods hardly use prior information of the segmentation target and their performance is far from learning from human annotations. 
\begin{figure*}[t]
\centerline{\includegraphics[width=2.00\columnwidth]{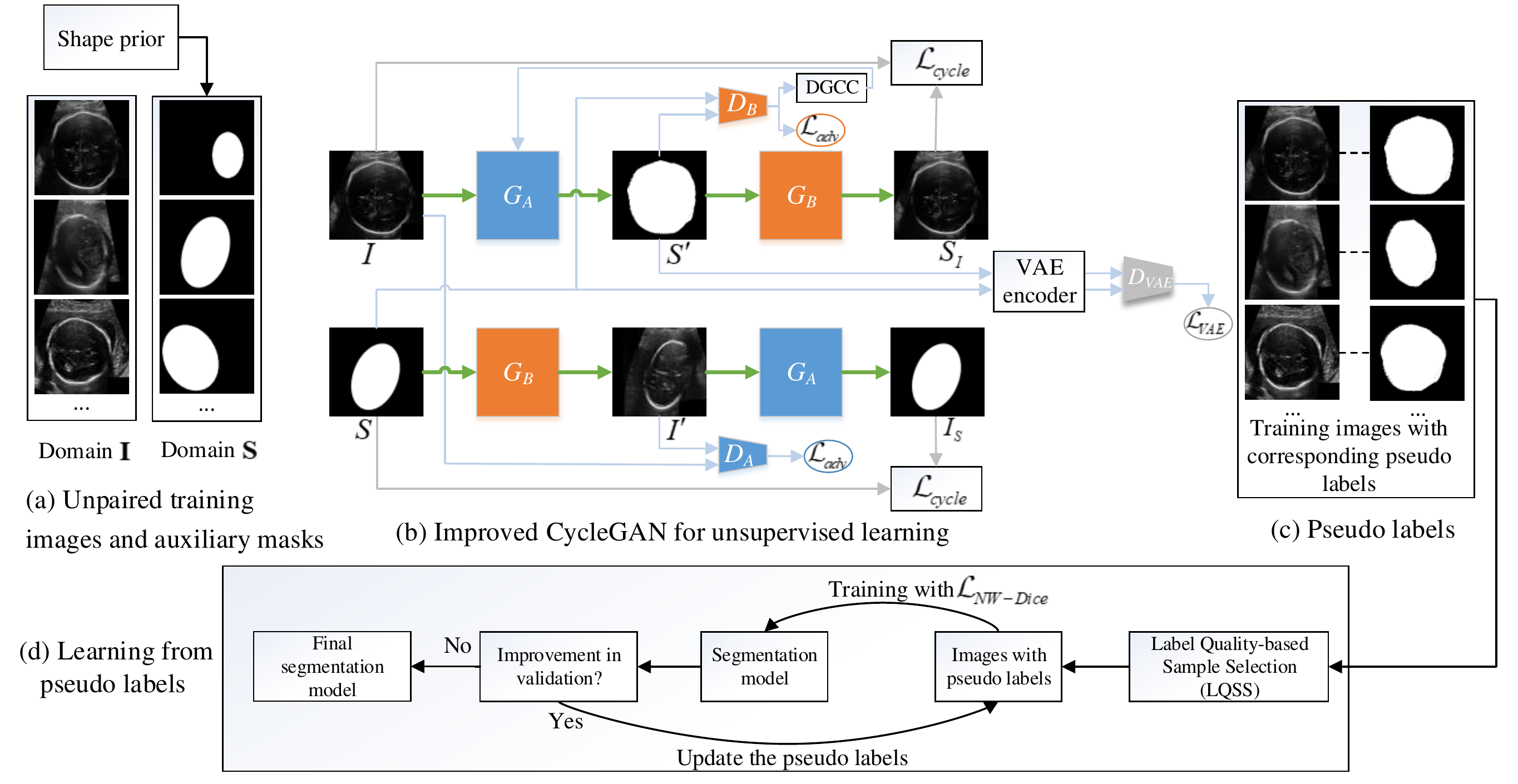}}
\caption{An overview of our proposed annotation-efficient deep learning method for medical image segmentation. a) We use a set of auxiliary masks (e.g., obtained from a shape prior model in fetal head segmentation) that are unpaired with training images for training. (b) An improved CycleGAN learns from the unpaired images and auxiliary masks to obtain the pseudo label corresponding to each training image, where a VAE-based discriminator and a DGCC module are proposed for better performance. (c) The pseudo labels for the training set. (d) A noise-robust iterative learning method to train the final segmentation model using the pseudo labels.}
\label{fig:fig1}
\end{figure*}
\subsubsection{Shape Constraint for Segmentation}
Shape models have been widely used to improve the robustness of segmentation methods~\cite{heimann2009statistical}. For example, spherical priors like SPHARM~\cite{styner2006framework} was proposed for brain structure analysis. Sparse shape composition~\cite{wang2015homotopy} was proposed to model complex shape structures with dictionary learning. Recently, Safar et al.~\cite{safar2015learning} proposed to learn shape priors for object segmentation via neural networks. Oktay et al.~\cite{oktay2017anatomically} encouraged models to follow the global anatomical properties of the underlying anatomy via learnt non-linear representations of the shape. Some other examples of employing shape models include Voxelmorph~\cite{balakrishnan2019voxelmorph} for registration and DeepSSM~\cite{bhalodia2018deepssm} for characterization and classification. Differently from these works, we employ VAE and adversarial learning to add an implicit high-level shape constraint on the segmentation output so that it follows the distribution of our set of auxiliary masks unpaired with training images.
\subsubsection{Learning from Noisy Labels}
Learning from noisy labels has been increasingly investigated recently due to the challenges to obtain high-quality annotations, and many existing works focus on image classification tasks~\cite{ghosh2017robust,xue2019robust}. For example, some novel loss functions, such as the Mean Absolute Error (MAE)~\cite{ghosh2017robust}, Generalized Cross Entropy~\cite{zhang2018generalized} and noise-robust Dice loss~\cite{9109297}, have been proposed to deal with noisy labels. Rusiecki et al.~\cite{rusiecki2019trimmed} proposed a trimmed cross entropy loss to exclude samples with large training errors. For medical image segmentation with noisy labels, Zhu et al.~\cite{zhu2019pick} proposed a strategy to evaluate the relative quality of training labels and thus only the good ones are used to tune the network parameter. Mirikharaji et al.~\cite{mirikharaji2019learning} assigned lower weights to pixels with abnormal loss gradient direction. However, these methods require a set of clean labels for training. Karimi et al.~\cite{karimi2019deep} used an iterative label update method to deal with simulated noisy labels, but its effectiveness on real noisy labels has not been investigated.



\section{Method}

Fig.~\ref{fig:fig1} shows an overview of our annotation-efficient learning method for segmentation, which avoids annotations for training images by learning from a set of auxiliary masks that are either generated from a parametric shape model or obtained from a publicly available dataset. It consists of two main stages: pseudo label generation based on shape constraints contained by auxiliary masks and noise-robust learning from pseudo labels. Both stages are critical for our framework, as the first stage is important for obtaining  high-quality pseudo labels, and the second stage is important for dealing with noise in pseudo labels for the final segmentation model. 

With the help of auxiliary masks that are unpaired with training images, we first use a generator to translate a medical image to its corresponding pseudo label based on an improved CycleGAN~\cite{zhu2017unpaired} framework that introduces implicit high-level shape constraints through adversarial learning.
We propose a VAE-based discriminator and a DGCC module which calibrates the pseudo label generator using the discriminator’s feedback for better pseudo labels. Then, we learn from the noisy pseudo labels to obtain the final segmentation model, and propose a noise-robust iterative training method based on a noise-weighted Dice Loss and Label Quality-based Sample Selection (LQSS) module to overcome the effect of noise and obtain high-performance segmentation model.

\subsection{Learning without Image-Annotation Pairs}
Let $\mathbf{I}$ and $\mathbf{S}$ represent the medical image domain and the segmentation mask domain, respectively. Differently from standard CNN-based image segmentation methods~\cite{ronneberger2015u,milletari2016v} that require samples from $\mathbf{S}$ to be manually provided so that they are paired with images from $\mathbf{I}$, we learn from two unpaired sets from $\mathbf{I}$ and $\mathbf{S}$, as it is more efficient to generate or collect a set of auxiliary masks from third-party sources rather than annotating the training images from $\mathbf{I}$.

First, considering that the segmentation mask in some applications has a strong shape prior (e.g., the fetal head), we take advantages of a shape model to generate a set of random samples from the segmentation mask domain. Specifically, for our tasks of fetal head and optic disc segmentation, the segmentation target looks like an ellipse. We therefore generate random ellipses in 2D space to simulate samples from domain $\mathbf{S}$. To make the shape of ellipses close to that of real segmentation target, we constrain the size, aspect ratio and orientation based on the prior distribution of corresponding values of the real target according to~\cite{campbell1977ultrasound,hadlock1982fetal}. For fetal head, we took the minor axis from 25 mm to 105 mm, aspect ratio from 1.2 to 1.8 and orientation from 0 to 2$\pi$. Then the generated ellipses are rasterized into binary images according to the pixel size of training images. Note that the position and shape of such a random mask does not correspond to any real training images, i.e., we obtain unpaired training images and random masks. Fig.~\ref{fig:fig1}(a) shows some examples of our generated random masks for the fetal head. 

Second, for a more complex segmentation structure that is hard to model (e.g., the lung and the liver), we can directly use a set of samples from the mask domain (unpaired to training images) for training when such auxiliary mask samples are available from other sources, e.g., public datasets. Note that once a set of auxiliary masks from domain $\mathbf{S}$ that are unpaired with training images have been obtained, the following training process is the same for these two situations. 
\subsection{Generating Pseudo Labels for Training Images}
\subsubsection{Cycle-consistent Adversarial Training}
With the auxiliary masks that are unpaired with our unannotated training images, we take advantage of their high-level shape information through adversarial training to constrain a generator $G_A$ so that $G_A$ generates pseudo labels for training images that have the same shape distribution as the auxiliary masks.
As shown in Fig.~\ref{fig:fig1}(b), given a medical image $I$ from domain $\mathbf{I}$ and an auxiliary mask $S$ from domain $\mathbf{S}$, we use a pseudo label generator $G_A$ to translate $I$ into a binary mask $S'= G_A(I)$ (i.e., pseudo label) corresponding to $I$, and $S'$ is translated back to a medical image $I_S = G_B(S')$ by the image generator $G_B$. On the contrary, an auxiliary mask sample $S$ from domain $\mathbf{S}$ is translated by $G_B$ into a pseudo medical image $I' = G_B(S)$ corresponding to $S$, and $I'$ is translated back to a binary mask $S_I = G_A(I')$. A cycle consistency loss which prevents the generators from producing a result that is irrelevant to the input between $I$ and $I_S$ (between $S$ and $S_I$, similarly) is calculated as:
\begin{figure*}[!t]
\centerline{\includegraphics[width=1.60\columnwidth]{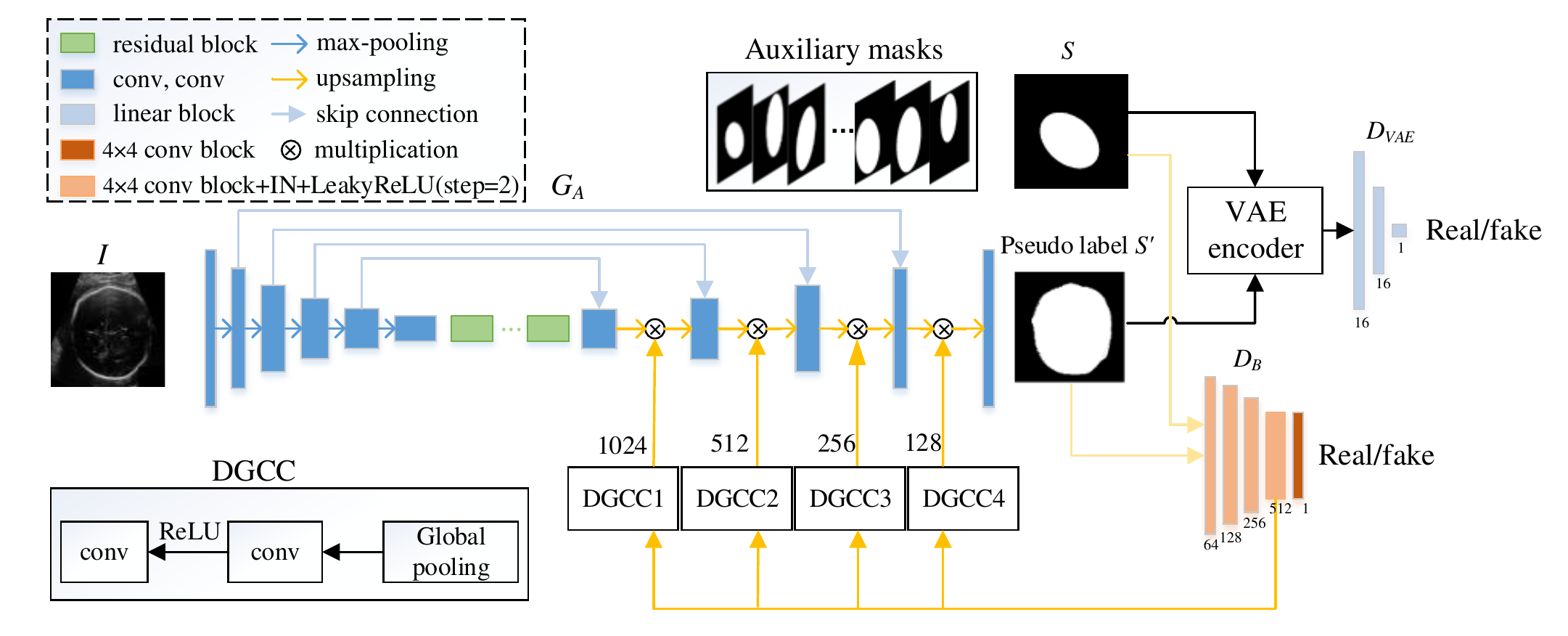}}
\caption{Illustration of our pseudo label generator $G_A$ that is calibrated by a feedback from the discriminator $D_B$ through our DGCC module. Another discriminator $D_{V\!AE}$ is introduced to assess the quality of generated pseudo labels in a high-level compact representation.}
\label{fig:fig2}
\end{figure*}
\begin{equation}
\mathcal{L}_{cycle}(G_A,G_B,\mathbf{I})=\mathbb{E}_{I\sim p_{data}(\mathbf{I})}\Big[||I-G_B(G_A(I))||_1\Big] \label{eq}
\end{equation}
where $p_{data}(\mathbf{I})$ is the distribution of domain $\mathbf{I}$. An adversarial loss~\cite{zhu2017unpaired,8653423} is used to encourage $S'$ to match the distribution of $S$:
\begin{equation}
\begin{split}
\mathcal{L}_{adv}(G_A,D_B)=\mathbb{E}_{I\sim p_{data}(\mathbf{I})}\Big[(1-D_B(G_A(I)))^2\Big]+\\
\mathbb{E}_{S\sim p_{data}(\mathbf{S})}\Big[D_B(S)^2\Big]
\end{split}\label{eq}
\end{equation}
where $p_{data}(\mathbf{S})$ is the distribution of domain $\mathbf{S}$ and the $D_B$ is a patch-based discriminator that distinguishes each patch of its input as a real or fake patch from domain $\mathbf{S}$. Our adversarial loss is least square adversarial loss~\cite{mao2017least} proposed for overcoming vanishing gradient problem of the original GAN loss~\cite{goodfellow2014generative}. $D_B$ serves to evaluate the quality of the pseudo label $S'$. Similarly, we use another discriminator $D_A$ to distinguish its input as a real or fake medical image. The original discriminator in~\cite{goodfellow2014generative} outputs a single scalar  that only indicates whether the input mask is real or fake as a whole, without giving details of non-local regions. In contrast, a patch-based discriminator~\cite{isola2017image} can better indicate the quality of subregions in the generator's output. Therefore, $D_A$ and $D_B$ are implemented by patch-based discriminators in this paper.
\subsubsection{VAE-based Discriminator}
Unlike those of regular images, pixel values in the pseudo segmentation label are not complex and sparse, which can be converted into a more compact representation by a latent vector. For example, an ellipse-like mask can be well represented by a low-dimensional vector specifying the size, position and orientation of the ellipse. Distinguishing the compact low-dimensional latent vectors additionally has a potential to obtain better performance than distinguishing the raw pseudo segmentation labels in a very high dimension only. 
Therefore, we propose to convert binary masks $S$ and $S’$ to their latent vector representations $z_r$ and $z_f$ respectively using a VAE~\cite{kingma2013auto}. We then apply a discriminator $D_{V\!AE}$ with three linear layers and leaky ReLU to distinguish them. VAE~\cite{kingma2013auto} is an encoder-decoder network, where the encoder network maps an input to a low-dimensional latent vector, and the decoder network attempts to reconstruct the input. We regularize the encoder by forcing the latent vector to follow a Gaussian distribution with a mean of zero and a variance of one. 

As the role  of VAE is to convert a segmentation mask in the 2D image space into a compact representation by a latent vector, we pretrain the VAE with the auxiliary masks. For pre-training, it takes an auxiliary mask as input and its decoder reconstructs the auxiliary mask as output. A KL divergence loss and an L2 loss were combined with an Adam optimizer for training. After pre-training, we fix the VAE and employ its encoder to obtain a compact representation of an input segmentation mask, which is sent into our $D_{V\!AE}$. The adversarial loss for $D_{V\!AE}$ can be written as:
\begin{equation}
\begin{split}
\mathcal{L}_{V\!AE}(z_f,z_r)=\mathbb{E}_{z_f\sim p_{data}(Z_F)}[(1-D_{V\!AE}(z_f))^2]+\\
\mathbb{E}_{z_r\sim p_{data}(Z_R)}[D_{V\!AE}(z_r)^2]
\end{split}\label{eq}
\end{equation}
where $Z_F$, $Z_R$ are sets of $z_f$, $z_r$, respectively. $p_{data}(Z_F)$, $p_{data}(Z_R)$ are the distributions of $Z_F$, $Z_R$, respectively. 

The overall loss of our method is summarized as:
\begin{equation}
\begin{split}
G_A^*,G_B^*=arg \mathop{min}\limits_{G_A,G_B} \mathop{max}\limits_{D_A,D_B,D_{V\!AE}} \lambda_{V\!AE}L_{V\!AE}(z_r,z_f)  \\ + \mathcal\lambda_{adv}[\mathcal{L}_{adv}(G_A,D_B)+
\mathcal{L}_{adv}(G_B,D_A)] \\ +
\mathcal\lambda_{cycle}[\mathcal{L}_{cycle}(G_A,G_B,\mathbf{I}) +{L}_{cycle}(G_A,G_B,\mathbf{S})]
\end{split}\label{eq4}
\end{equation}
where $\lambda_{V\!AE}$, $\lambda_{adv}$ and $\lambda_{cycle}$ control the relative weights of the three terms, respectively.
\subsubsection{Discriminator-guided Generator Channel Calibration}
In a standard GAN~\cite{zhu2017unpaired}, the discriminator gives a feedback to the generator by the loss function with back-propagation, which can only be used for training, i.e., indirect and implicit feedback. As the patch-based discriminator $D_B$ indicates whether a patch of pseudo label $S^{'}$ is real or fake and also learns the typical representation features~\cite{radford2015unsupervised}, the feature map of $D_B$ has a potential to explicitly guide $G_A$ to get better results. Besides, as the discriminator easily outperforms the generator, the generator $G_A$ could learn better and faster when calibrated by the feature map of $D_B$. Therefore, we propose a Discriminator-guided Generator Channel Calibration (DGCC) module to boost the performance of $G_A$. As shown in Fig.~\ref{fig:fig1}(b), we use four DGCC modules to calibrate the features of $G_A$ at four scales, respectively. 

In our DGCC, leveraging the discriminator’s feedback leads to recurrent loop connections. Let $T$ represent the total turn number in the loop connections, as shown in Fig.~\ref{fig:fig2}. At turn 1, the generator $G_A$ does not have feedback from the discriminator $D_B$. At the following few turns, we take $D_B$'s embedding feature map right before the output layer at turn $t$ as our feedback information:
\begin{equation}
    h_{t}=D_B(S'_{t})
\label{eq1}
\end{equation}
where $h_t \in \mathbb{R}^{C \times h \times w}$, and $C$, $h$, $w$ are the channel number, height and width of the embedding feature map of $D_B$, respectively. $S'_{t}$ is the $S'$ at the turn $t$. Then we apply a global average pooling ($P$) to obtain the average feature for each channel and use Sequeeze-and-Excitation (SE)~\cite{hu2018squeeze} consisting of two convolution layers to obtain an attention coefficient vector $\beta^s_t$ with a length of $C$ that equals to the channel number of $G_A$'s feature map at scale $s$ of turn $t$:
\begin{equation}
    \beta^s_{t}  = W_2 \cdot \delta (W_{1}\cdot P(h_{t}))
\label{eq}
\end{equation}
where $\delta$ refers to ReLU, $W_1\in \mathbb{R}^{\frac{C}{r}\times{C}}$ and $W_2\in \mathbb{R}^{{C}\times\frac{C}{r}}$ are $1 \times1$ convolution layers and $r$ is set to 4 according to common practice~\cite{hu2018squeeze}. Let $u^s_{t+1}$ be a certain feature map at scale $s$ in the decoder of $G_A$ before calibration at turn $t+1$ of the recurrence, the corresponding calibrated feature map $\hat{u}^s_{t+1}$ at turn $t+1$ of scale $s$ is:
\begin{equation}
    \hat{u}^s_{t+1}= \beta^s_t  u^s_{t+1} + u^s_{t+1}
\label{eq}
\end{equation}
where a residual connection is used to facilitate the training. The new mask ${S}^{'}_{t+1}$ obtained by $G_A$ is:
\begin{equation}
    {S}^{'}_{t+1}=G_A(I, \beta^s_t)
\label{eq}
\end{equation}
For testing, we take ${S}^{'}_T$ at turn $T$ of the recurrent connection as the pseudo segmentation label obtained by $G_A$.

\subsection{Learning from Noisy Pseudo Labels}
After training with the unpaired images and random masks described above, $G_A$ can be used to predict a corresponding pseudo segmentation label for each training image. With these pseudo labels, one may use a supervised training pipeline to train a  segmentation model such as U-Net~\cite{ronneberger2015u} with the standard Dice loss. However, differently from the labels of standard supervised training, our pseudo labels are noisy and not accurate. To address this problem, we propose a two-step framework that learns from noisy pseudo labels given by $G_A$, as shown in Fig.~\ref{fig:fig1}(d).

In the first step, we propose a Label Quality-based Sample Selection (LQSS) method  to automatically reject pseudo labels with low quality and only keep high-quality pseudo labels. According to GAN~\cite{goodfellow2014generative}, a well-trained discriminator $D_B$ can indicate whether its input is a real or fake sample from the segmentation mask domain. Note that our patch-based discriminator $D_B$'s output is an $N \times N$ matrix where each element indicates the quality of the corresponding patch. For a training image $I_i$, We take the average value of that matrix as an image-level quality score $R_i$ of the corresponding pseudo segmentation label $Y_i$. The training set $\mathcal{T}$ with pseudo labels can be represented as $\mathcal{T} = \{(I_1, Y_1, R_1), (I_2, Y_2, R_2), ... (I_N, Y_N, R_N)\}$. The training set after LQSS is:
\begin{equation}
    \mathcal{T'} = \{(I_i, Y_i, R_i) \, | \, (I_i, Y_i, R_i) \in \mathcal{T} \, \text{and} \, R_i  < \alpha\}
\label{eq}
\end{equation}
where $\alpha$ is a threshold value for the pseudo label's quality score and is set as 75 percentile of  $S_i$ in $\mathcal{T}$.

In the second step, from the selected images with high-quality pseudo labels, we use an iterative training procedure with $K$  rounds, and each round consists of 1) updating the segmentation model by learning from the pseudo labels and 2) predicting new pseudo labels for training images using the current segmentation model, which is illustrated in Fig.~\ref{fig:fig1}(d). The round stops when there is no improvement of segmentation performance on the validation set. During the segmentation model update step, considering  that some pixels in pseudo labels are noisy and even outliers, which would seriously corrupt the segmentation model, we propose to weight each pixel based on the estimated noise level. As samples with wrong labels are likely to cause high loss values~\cite{rusiecki2019trimmed}, we assign lower weights to pixels with large training error to reduce the effect of potentially noisy labels. The noise-weighted Dice loss is formulated as:
\begin{equation}
    \mathcal{L}_{N\!W-Dice}=1.0 - \sum\frac{2\sum_{i}w_{i}p_{i}g_{i}+\epsilon}{\sum_{i}w_{i}(p_{i}+g_{i})+\epsilon}
\label{eq}
\end{equation}
where $\epsilon=10^{-5}$ is a small number for numerical stability. $p_{i}$ and $g_{i}$ are the foreground probability for pixel $i$ in the segmentation result and the pseudo label, respectively. The weight $w_{i}$ is defined as:
\begin{equation}
    {w}_{i}=1-\mid{p}_{i}-{g}_{i}\mid
\label{eq}
\end{equation}
\section{EXPERIMENTS AND RESULTS}
We validated our proposed annotation-efficient framework for segmentation in two situations. 1) Easy-to-model structures like the optic disc in retinal fundus images and the fetal head in ultrasound images, where a parametric shape model is used to obtain the auxiliary masks.  2) Complex structure like lung in X-Ray images and liver in CT images, where auxiliary masks are obtained from public datasets. For quantitative evaluation of segmentation performance, we measured Dice, Average Symmetric Surface Distance (ASSD) between segmentation results and the ground truth.

\subsection{Implementation Details}
We implemented our networks in PyTorch with two NVIDIA GTX1080 Ti GPUs. The architecture of our generator is a variant of U-Net~\cite{ronneberger2015u} where we added six residual blocks to the bottleneck for higher feature representation ability. We set the channel number in the first block as 64, and it is doubled after each down-sampling layer in the encoder, as shown in Fig.~\ref{fig:fig2}. $D_A$ and $D_B$ were implemented by 70 $\times$ 70 PatchGANs~\cite{isola2017image}. $h_t$ in Eq.~\eqref{eq1} has a channel number of 512, and the channel numbers of the calibration coefficient $\beta^s_{t}$ are set to 1024, 512, 256 and 128 for our four DGCC modules, respectively, and they are equal to the corresponding channel numbers in the decoder of $G_A$, as shown in Fig.~\ref{fig:fig2}. The latent vector length for our VAE was set to 32. For training, we used Adam optimizer with an initial learning rate of 5 $\times$ $10^{-6}$ in the first 50 epochs and it was linearly decayed to 0 in the following 100 epochs. As Eq~\eqref{eq4} for pseudo label generation is an extension of CycleGAN framework, we set $\lambda_{cycle}=10$ and $\lambda_{adv}=1$ according to CycleGAN~\cite{zhu2017unpaired}. For our newly introduced $\lambda_{V\!AE}$, we used grid search to find its optimal value (i.e., 1.0) according to the validation set. 

\begin{figure}[!t]
\centerline{\includegraphics[width=1.00\columnwidth]{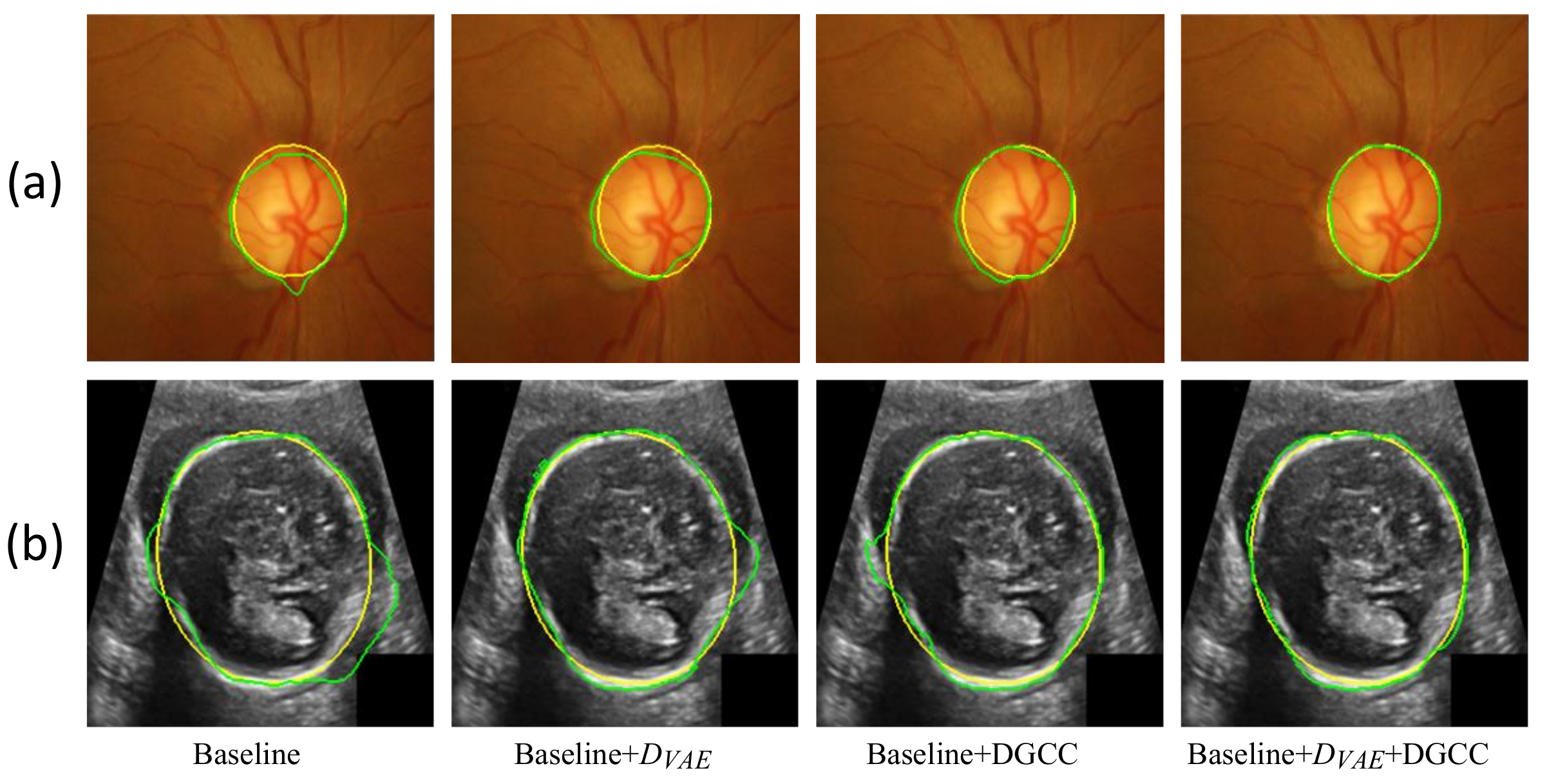}}
\caption{Visual comparison of results of pseudo label generator $G_A$ with different training methods. Green and yellow curves show the output of $G_A$ and the ground truth, respectively. (a) and (b) show results for optic disc and fetal head, respectively.}
\label{fig:fig3}
\end{figure}
\begin{table}
	\centering
	\setlength{\tabcolsep}{5pt}
    \scriptsize
	\caption{The effect of $D_{V\!AE}$ on the pseudo label generator $G_A$ for optic disc and fetal head segmentation.}
	\label{tab:vaeo}
	\begin{tabular}{lcccc}
	\toprule
	\multirow{2}{*}{Methods} &\multicolumn{2}{c}{Optic Disc} &\multicolumn{2}{c}{Fetal head} \\ \cline{2-5}
	&Dice & ASSD (pixel) &Dice & ASSD (pixel)\\ \hline
    baseline & 0.909$\pm$0.042 & 4.623$\pm$2.306 & 0.904$\pm$0.078 & 9.130$\pm$5.977\\
    $D_{V\!AE}$(w/o $D_{B}$) & 0.906$\pm$0.043 & 6.558$\pm$3.828 & 0.896$\pm$0.064 & 8.965$\pm$6.019\\
    $D_{V\!AE}$(beta) & 0.917$\pm$0.043 & 4.373$\pm$3.240 & 0.916$\pm$0.097 & 7.085$\pm$8.202\\
    $D_{V\!AE}$ & \textbf{0.918$\pm$0.038} & \textbf{3.770$\pm$1.957} & \textbf{0.918$\pm$0.085} & \textbf{6.890$\pm$5.497}\\
	\bottomrule
	\end{tabular}
\end{table}
\begin{table}
	\centering
	\setlength{\tabcolsep}{5pt}
	\scriptsize
	\caption{Effect of latent vector length of VAE on the output of $G_A$ for optic disc and fetal head segmentation, based on the validation set.}
	\label{tab:vaev}
	\begin{tabular}{lcccc}
	\toprule
	\multirow{2}{*}{Length}  &\multicolumn{2}{c}{Optic Disc} &\multicolumn{2}{c}{Fetal head} \\ \cline{2-5}
	&Dice & ASSD (pixel) &Dice & ASSD (pixel)\\ \hline

    16  &0.906$\pm$0.050 & 5.694$\pm$6.237 & 0.922$\pm$0.077 & 5.604$\pm$3.888\\
    32  & \textbf{0.908$\pm$0.051} & \textbf{4.675$\pm$4.630} & \textbf{0.928$\pm$0.085} & \textbf{5.392$\pm$5.485}\\
    64  & 0.905$\pm$0.068 & 5.255$\pm$2.621 & 0.928$\pm$0.087 & 5.474$\pm$5.298\\
	\bottomrule
	\end{tabular}
\end{table}
\subsection{Segmentation of Structure with Shape Prior Models}
We first apply our annotation-efficient segmentation framework to structures with strong shape priors, where a parametric shape model can be used to obtain a set of auxiliary masks required by our method. For the experiment, we consider the segmentation of Optic Disc (OD) from fundus images and fetal head from ultrasound, where both objects can be modeled as ellipses. Segmentation of these structures are important for  ophthalmic disease diagnosis~\cite{zheng2013optic,joshi2011optic,yin2019pm} and fetal growth assessment~\cite{van2018automated}.

\subsubsection{Data}
For optic disc segmentation, we utilized the Digital Retinal Images for Optic Nerve Head (optic disc and cup) Segmentation Database (DRIONS-DB)~\cite{carmona2008identification} and retinal image dataset for optic nerve head segmentation (Drishti-gs)~\cite{sivaswamy2014drishti,sivaswamy2015comprehensive}. DRIONS-DB consists of 110 colour digital retinal images with a size of 600 $\times$ 400. Drishti-gs consists of 101 colour digital retinal images with a varying image size. Each image in these two datasets had annotations of optic disc by two and four experts, respectively. We averaged these multiple segmentation contours along the radical direction for a given image as the ground truth. As these two datasets are relatively small, we merged them into a single dataset for experiments. As the segmentation target is relatively small and located near the center of the image, we cropped the images at the center to 60$\%$ of the original size. For fetal head segmentation, we used the HC18 dataset\footnote{http://doi.org/10.5281/zenodo.1322001}~\cite{van2018automated} containing 999 2D ultrasound images of the fetal head in the standard plane for experiment. The ultrasound images were acquired from 551 pregnant women in all trimesters of the pregnancy, and all the cases did not exhibit any growth abnormalities. The size of each 2D ultrasound image was 800 $\times$ 540 with a pixel size ranging from 0.052 mm to 0.326 mm. For these two applications, the images were randomly split into 70$\%$, 10$\%$, 20$\%$ for training, validation and testing, respectively. We abandoned the ground truth of the training set for our annotation-efficient learning. Each image was resized to $288 \times 288$, randomly cropped to 256 $\times$ 256, and the intensity was normalized into the range of [-1, 1].
\begin{figure}[!t]
\centerline{\includegraphics[width=1.00\columnwidth]{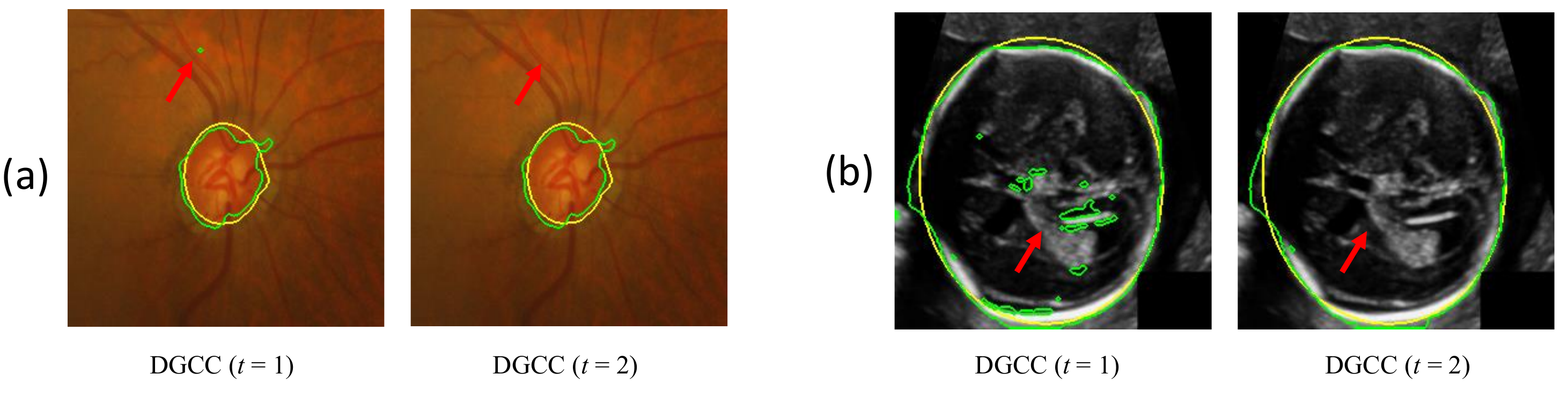}}
\caption{Visual comparison between outputs of pseudo label generator $G_A$ before ($t=1$) and after ($t=2$)  channel calibration by our DGCC module. (a) and (b) show results for optic disc and fetal head, respectively. Green and yellow curves are results of $G_A$ and the ground truth, respectively. }
\label{fig:fig4}
\end{figure}
\begin{figure}[!t]
\centerline{\includegraphics[width=1.00\columnwidth]{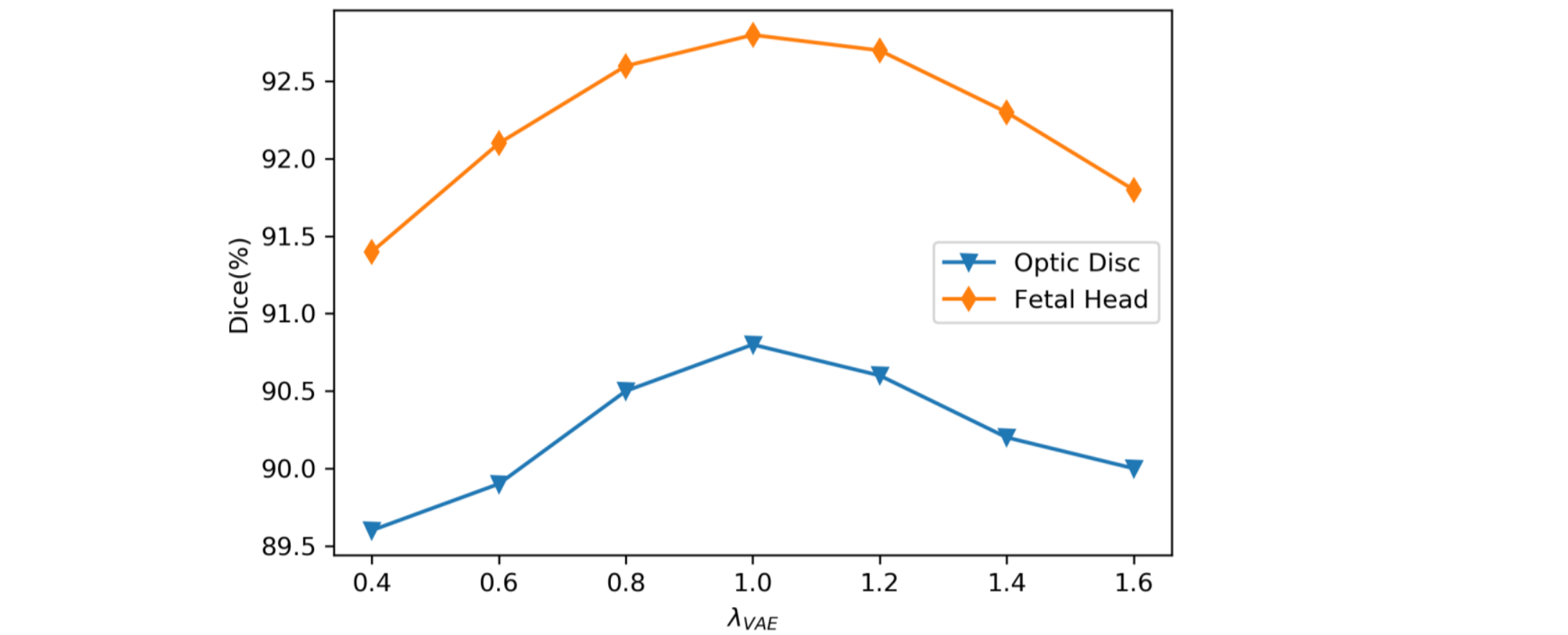}}
\caption{Effect of $\lambda_{V\!AE}$ on the performance of $G_A$ for our validation set.}
\label{fig:figh}
\end{figure}
\begin{table}
	\centering
	\setlength{\tabcolsep}{5pt}
	\scriptsize
	\caption{Quantitative evaluation of the effect of DGCC on $G_A$ for optic disc and fetal head segmentation.}
	\label{tab:feedbacko}
    \begin{tabular}{lcccc}
	\toprule
	\multirow{2}{*}{Methods} &\multicolumn{2}{c}{Optic Disc} &\multicolumn{2}{c}{Fetal head} \\ \cline{2-5}
	&Dice & ASSD (pixel) &Dice & ASSD (pixel)\\ \hline
    Baseline & 0.909$\pm$0.042 & 4.623$\pm$2.306  & 0.904$\pm$0.078 & 9.130$\pm$5.977 \\
	+DGCC(l) & 0.917$\pm$0.040 & 4.003$\pm$2.525& 0.913$\pm$0.130 & 7.083$\pm$8.488 \\
    +DGCC(h) & 0.914$\pm$0.044 & 4.375$\pm$4.231 & 0.908$\pm$0.109 & 8.906$\pm$7.713\\
	+DGCC($t=1$) & 0.921$\pm$0.032 &3.794$\pm$2.246 & 0.917$\pm$0.097 & 8.391$\pm$7.386 \\   
	+DGCC & 0.922$\pm$0.032 & 3.774$\pm$2.262  & 0.921$\pm$0.093 & 6.800$\pm$7.061 \\   
	+$D_{V\!AE}$+DGCC & \textbf{0.937$\pm$0.036} & \textbf{3.174$\pm$2.468} & \textbf{0.937$\pm$0.086} & \textbf{5.331$\pm$6.985}\\
	\bottomrule
	\end{tabular}
\end{table}

\subsubsection{Effectiveness of VAE-based Discriminator and DGCC}
We first evaluate the effectiveness of our pseudo label generation method.
For an ablation study of our $D_{V\!AE}$, we started with the baseline of training a CycleGAN~\cite{zhu2017unpaired} without $D_{V\!AE}$ and DGCC. Table~\ref{tab:vaev} shows quantitative evaluation results of the output of $G_A$ on the validation set with different latent vector length of the VAE. It can be observed that the best performance is achieved when the length of latent vector is 32. Fig~\ref{fig:figh} shows that the optimal value of hyper-parameter $\lambda_{V\!AE}$ is 1.0, and the performance of $G_A$ does not change much when $\lambda_{V\!AE}$ is around 1.0. We also compared our $D_{V\!AE}$ with three counterparts: 1) the baseline not using $D_{V\!AE}$, 2) our $D_{V\!AE}$ without $D_B$, 3) replacing the VAE with beta-VAE~\cite{higgins2017beta} where the latent vector length was 32, denoted as $D_{V\!AE}$(beta). Quantitative evaluation based on the testing set in Table~\ref{tab:vaeo} shows that our $D_{V\!AE}$ outperformed the counterparts, and compared with not using $D_{V\!AE}$, it improved the average Dice from 0.909 to 0.918 for optic disc segmentation and from 0.904 to 0.918 for fetal head segmentation, respectively. Fig.~\ref{fig:fig3} demonstrates the effectiveness of our VAE-based discriminator on the output of $G_A$. 

We further evaluated the effectiveness of our multi-scale DGCC module by ablation studies. We compared it with two variants: 1) DGCC(l) that only calibrates the low-resolution feature map obtained by the bottleneck of $G_A$; 2) DGCC(h) that only calibrates the high-resolution feature map before the last convolution block of $G_A$. The quantitative evaluation results of these variants combined with our baseline model are shown in Table~\ref{tab:feedbacko}. It can be observed that our multi-scale DGCC has a higher performance than DGCC(l) and DGCC(h), which demonstrates that multi-scale calibration performed better than single-scale calibration of the pseudo label generator $G_A$. We also compared the calibrated result at $t=2$ of our DGCC with the result at $t=1$ (i.e., before calibration) of our DGCC module, which is denoted as DGCC($t=1$). The quantitative results in Table~\ref{tab:feedbacko} and qualitative results in Fig.~\ref{fig:fig4} show that the calibration helps to reduce and even remove some noise in the output of $G_A$. In addition, Fig.~\ref{fig:fig3} and Table~\ref{tab:feedbacko} show that combining our $D_{V\!AE}$ with DGCC outperforms the other variants.
\begin{figure}[!t]
\centerline{\includegraphics[width=0.5\textwidth]{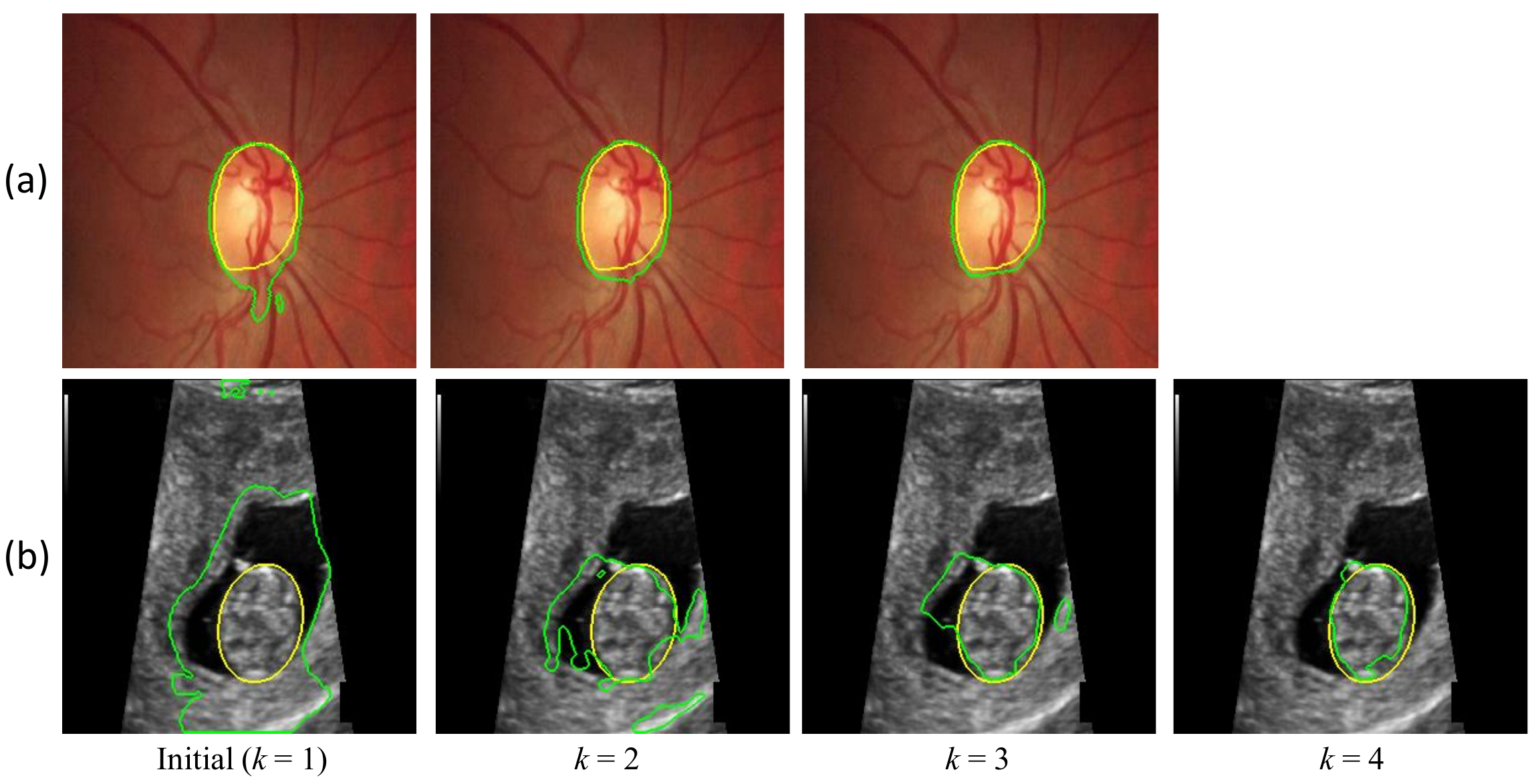}}
\caption{Visual comparison of pseudo labels (green curves) of training images at different rounds of our proposed iterative training method with noise-weighted Dice loss. (a) and (b) show results for optic disc and fetal head, respectively. Yellow curves are the ground truth labels.}
\label{fig:fig5}
\end{figure}
\begin{figure}[!t]
\centerline{\includegraphics[width=1.00\columnwidth]{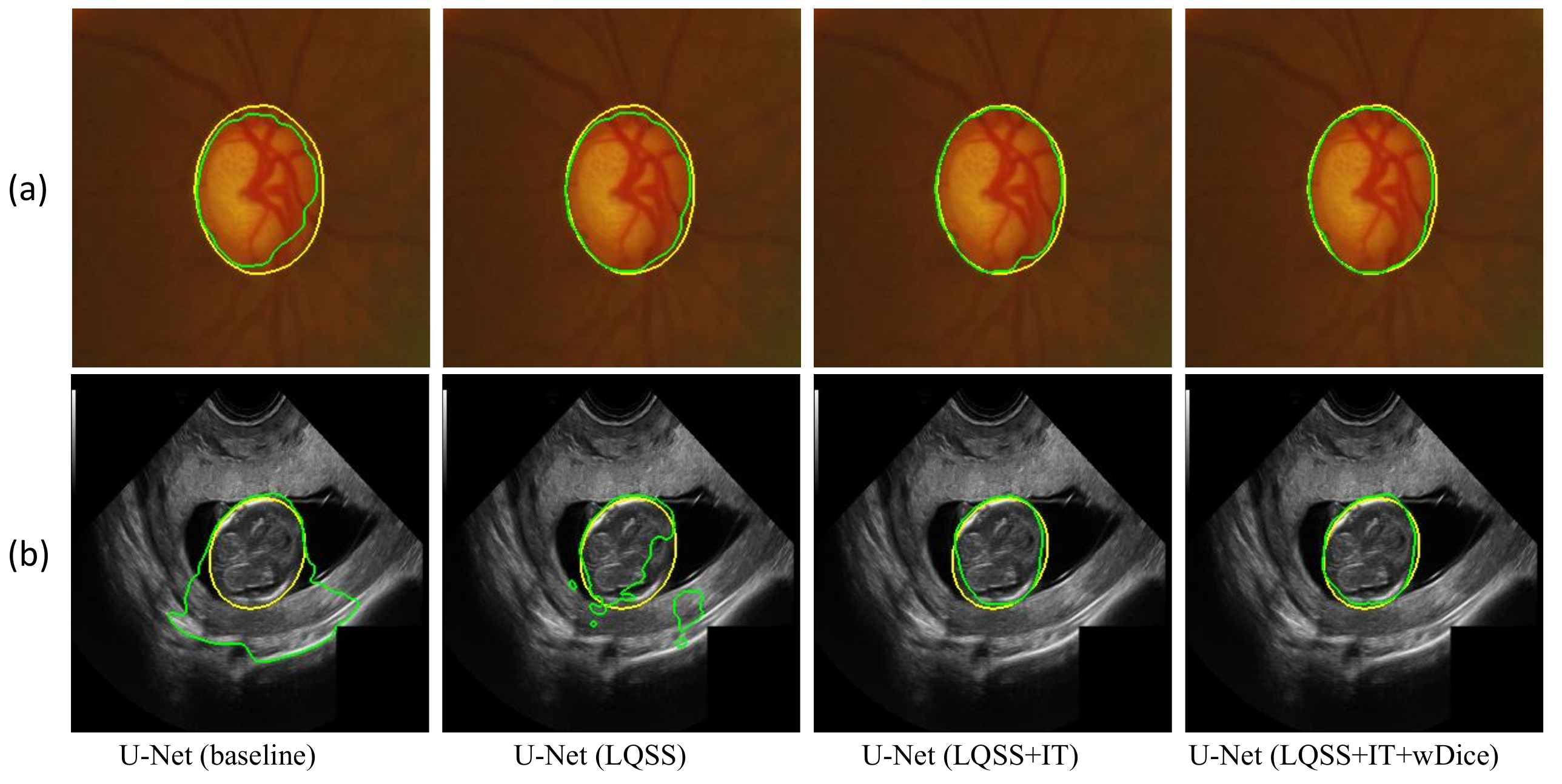}}
\caption{Visual comparison between different variants of our method for learning from noisy pseudo labels obtained by $G_A$. (a) and (b) show results for optic disc and fetal head, respectively. Green and yellow curves are segmentation results and the ground truth, respectively.}
\label{fig:fig6}
\end{figure}
\begin{figure}[!t]
\centerline{\includegraphics[width=1.00\columnwidth]{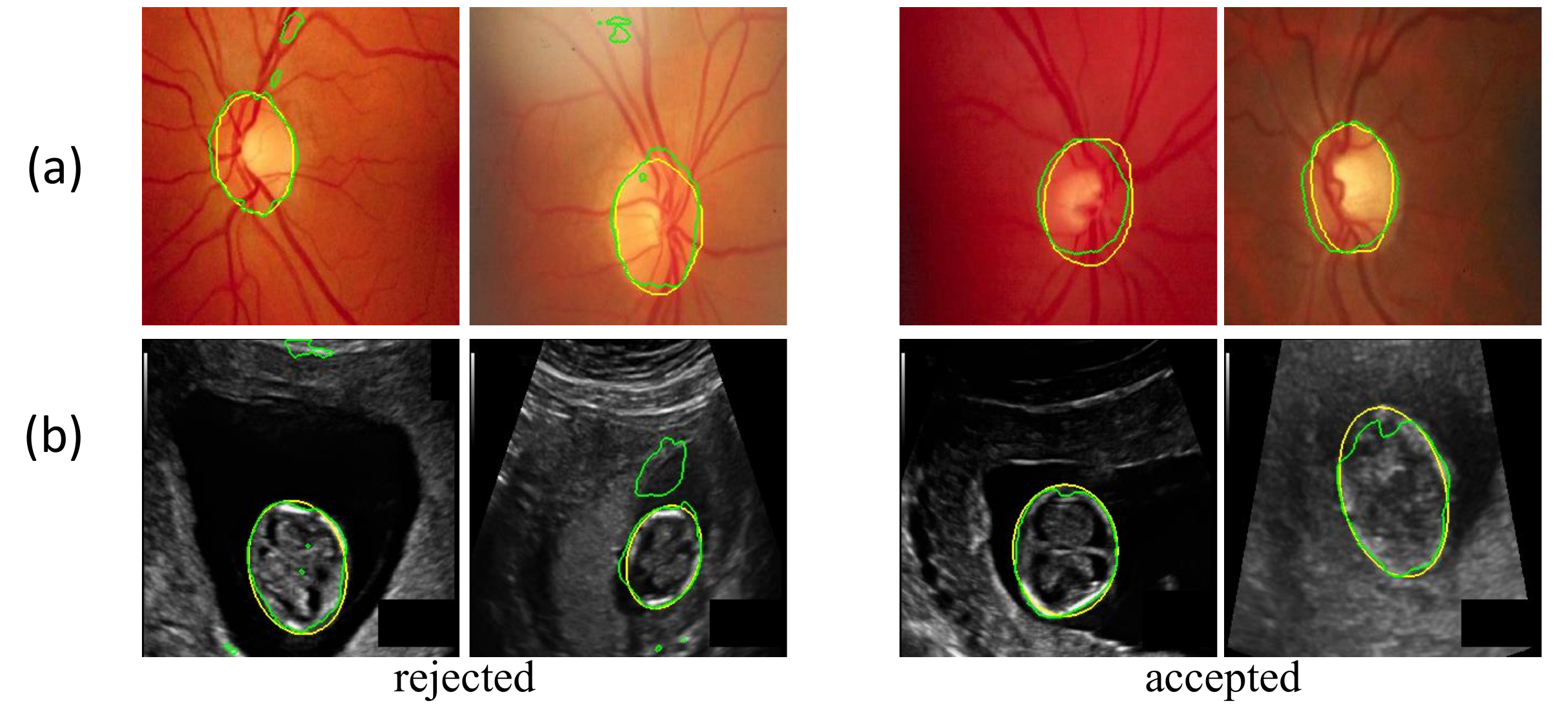}}
\caption{Examples of rejected and accepted samples by our LQSS. (a) and (b) show results for optic disc and fetal head, respectively.}
\label{fig:fig7}
\end{figure}
\begin{figure}[!t]
\centerline{\includegraphics[width=1.00\columnwidth]{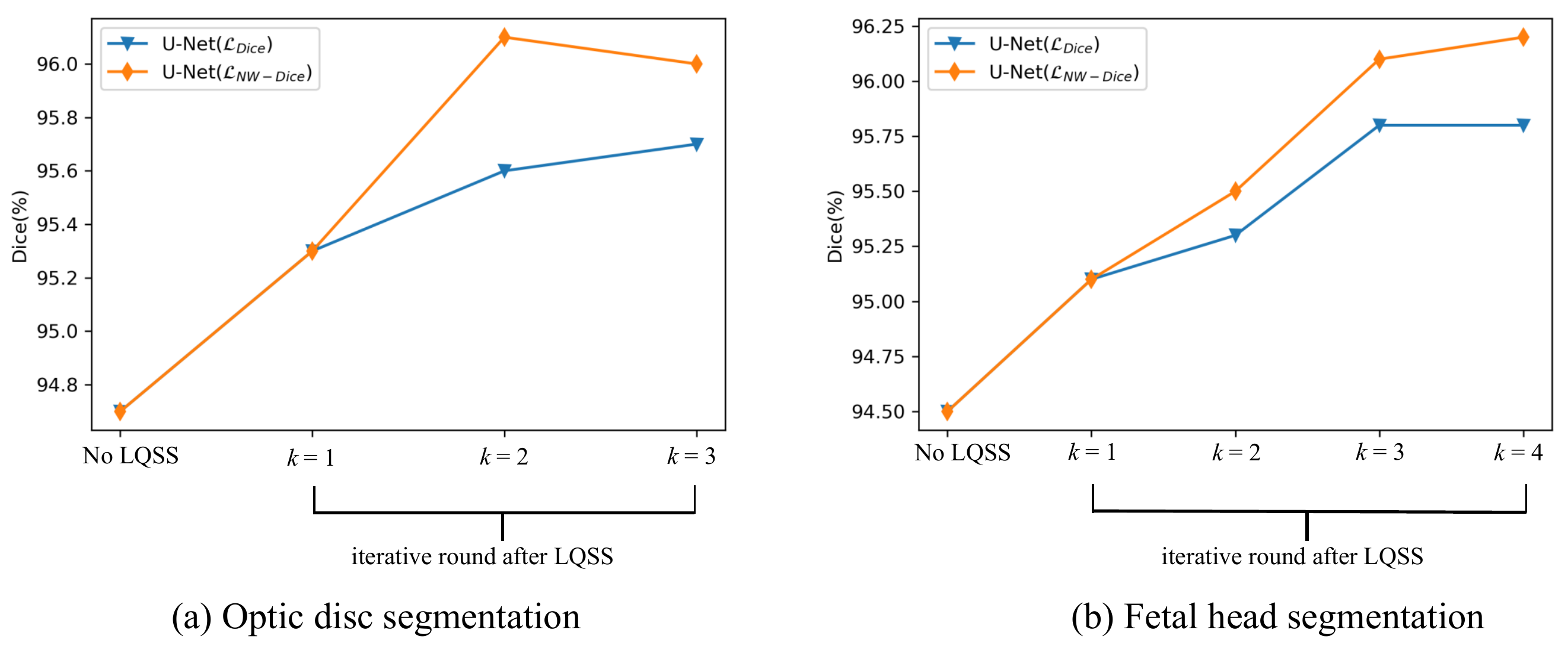}}
\caption{Effect of our proposed iterative training method for learning from pseudo labels obtained by $G_A$. $k$ represents the number of rounds. (a) and (b) show results for optic disc and fetal head, respectively.}
\label{fig:fig8}
\end{figure}
\begin{table}
	\centering
	\setlength{\tabcolsep}{0.5pt}
	\scriptsize
	\caption{Quantitative evaluation of different methods for optic disc and fetal head segmentation. Bold font highlights the best results obtained by our final segmentation model.  Results with no significant difference from U-Net (manual) are denoted by *, according to a paired t-test (p-value $>$ 0.05). $^\#$ denotes learning from pseudo labels obtained by our $G_A$. $^\square$ denotes learning from manual annotations. $^\triangle$ and $^\perp$ denote results based on exiting unsupervised and weakly supervised methods, respectively.  \cite{joshi2011optic} and~\cite{lu2019weakly} are only for optic disc, and \cite{perez2015automatic} is only for fetal head.}
	\label{tab:noiseo}
	\begin{tabular}{lcccc}
	\toprule
	\multirow{2}{*}{Methods} &\multicolumn{2}{c}{Optic Disc} &\multicolumn{2}{c}{Fetal head} \\ \cline{2-5}
	&Dice & ASSD (pixel) &Dice & ASSD (pixel)\\ \hline
    $^\#$U-Net (baseline) & 0.947$\pm$0.044 & 2.356$\pm$1.743 & 0.945$\pm$0.061 & 4.238$\pm$3.604\\
	$^\#$U-Net (MAE) & 0.945$\pm$0.044 & 2.406$\pm$1.657 & 0.945$\pm$0.059 & 4.313$\pm$3.490\\
	$^\#$U-Net ($\mathcal{L}_{gc}$)  & 0.905$\pm$0.116 & 3.869$\pm$2.775 & 0.934$\pm$0.097 & 5.635$\pm$6.097\\
	$^\#$U-Net (LQSS) & 0.953$\pm$0.024 & 2.214$\pm$1.010 & 0.951$\pm$0.041 & 4.137$\pm$3.487\\
	$^\#$U-Net (LQSS+IT) & 0.957$\pm$0.031 & 1.954$\pm$1.207 & 0.958$\pm$0.034 & 3.530$\pm$2.561\\ 
	$^\#$U-Net (LQSS+IT+wDice) & \textbf{0.961$\pm$0.018}* & \textbf{1.872$\pm$1.224}* & \textbf{0.962$\pm$0.026} & \textbf{3.155$\pm$2.153}\\ \hline
	$^\square$U-Net (manual) & 0.965$\pm$0.032 &1.580$\pm$1.228 & 0.973$\pm$0.018 &2.352$\pm$1.635\\
	$^\triangle$Joshi et al.~\cite{joshi2011optic} & 0.947$\pm$0.037 &2.321$\pm$1.603 & n/a & n/a\\
	$^\triangle$~\cite{joshi2011optic} + IT + wDice & 0.954$\pm$0.048 & 2.085$\pm$1.789 & n/a & n/a \\
	
	$^\triangle$Perez-Gonzalez et al.~\cite{perez2015automatic} & n/a& n/a & 0.804$\pm$0.179 & 16.929$\pm$13.967\\
	$^\triangle$~\cite{perez2015automatic} + IT + wDice & n/a & n/a & 0.881$\pm$0.144 & 9.077$\pm$8.739\\
	$^\triangle$Moriya et al.~\cite{moriya2019unsupervised} & 0.724$\pm$0.248 &13.817$\pm$14.003 & 0.523$\pm$0.136 & 35.934$\pm$8.35\\
	$^\perp$Kervadec et al.~\cite{kervadec2019constrained} & 0.887$\pm$0.062 &6.785$\pm$8.265 & 0.832$\pm$0.124 & 12.207$\pm$8.425\\
	$^\perp$Lu et al.~\cite{lu2019weakly} & 0.878$\pm$0.098 & 5.364$\pm$3.713 & n/a & n/a\\
	U-Net (baseline)$^{\circ}$ & 0.887$\pm$0.073 & 7.166$\pm$5.440 & 0.784$\pm$0.152 & 15.674$\pm$8.626\\
	\bottomrule
	\end{tabular}
\end{table}
\subsubsection{Results of Learning from Noisy Pseudo Labels}
To validate our noise-robust iterative method to learn from noisy pseudo labels obtained by $G_A$, we compared the following variants: 1) U-Net (baseline) that learns from the pseudo labels using a standard Dice loss without considering the existence of noise; 2) U-Net (MAE) that uses MAE loss~\cite{ghosh2017robust} for training; 3) U-Net ($\mathcal{L}_{gc}$) that uses generalized cross entropy loss~\cite{zhang2018generalized} for training; 4) U-Net trained with Dice loss from samples selected by our LQSS, which is referred to as U-Net (LQSS). These four methods only train the model once without iterative training, and were further compared with: 5) U-Net (LQSS + IT) that refers to U-Net (LQSS) followed by iterative training with Dice loss; and 6)  U-Net (LQSS + IT + wDice) that refers to U-Net (LQSS) followed by iterative training with our noise-weighted Dice loss. For the last two variants, the round number determined by the validation set was 3 and 4 for optic disc segmentation and fetal head segmentation, respectively. The quantitative evaluation results are shown in Table~\ref{tab:noiseo}, which shows that LQSS obtained better performance than the baseline, and using iterative training and noise-weighted Dice loss further improves the segmentation accuracy. Fig.~\ref{fig:fig7} shows that our LQSS is able to reject low-quality pseudo labels with some noise, e.g., over segmentation with false positives. Note that in Fig.~\ref{fig:fig7}(a), the second rejected case of has a higher contrast than the first accepted case, which shows our LQSS does not tend to only select easy samples. Fig.~\ref{fig:fig5} demonstrates the refinement of pseudo labels at different rounds of training stage. Fig.~\ref{fig:fig8} shows the performance at different rounds of our iterative method to learn from noisy pseudo labels obtained by $G_A$. It shows that the performance increased at the beginning and reached a plateau after two rounds for optic disc and three rounds for fetal head, and that noise-weighted Dice loss is better than Dice loss during the iterative training. We compared our ellipse-based shape prior with circle-based shape prior to obtain the pseudo labels, and they are denoted as U-Net (baseline) and U-Net (baseline)$^{\circ}$, respectively. Results in Table~\ref{tab:noiseo} show that modeling optic disc and fetal head as ellipses largely outperform modeling as circles. 

\subsubsection{Comparison with Existing Methods and Learning from Human Annotations}
\begin{figure}[!t]
\centerline{\includegraphics[width=1.00\columnwidth]{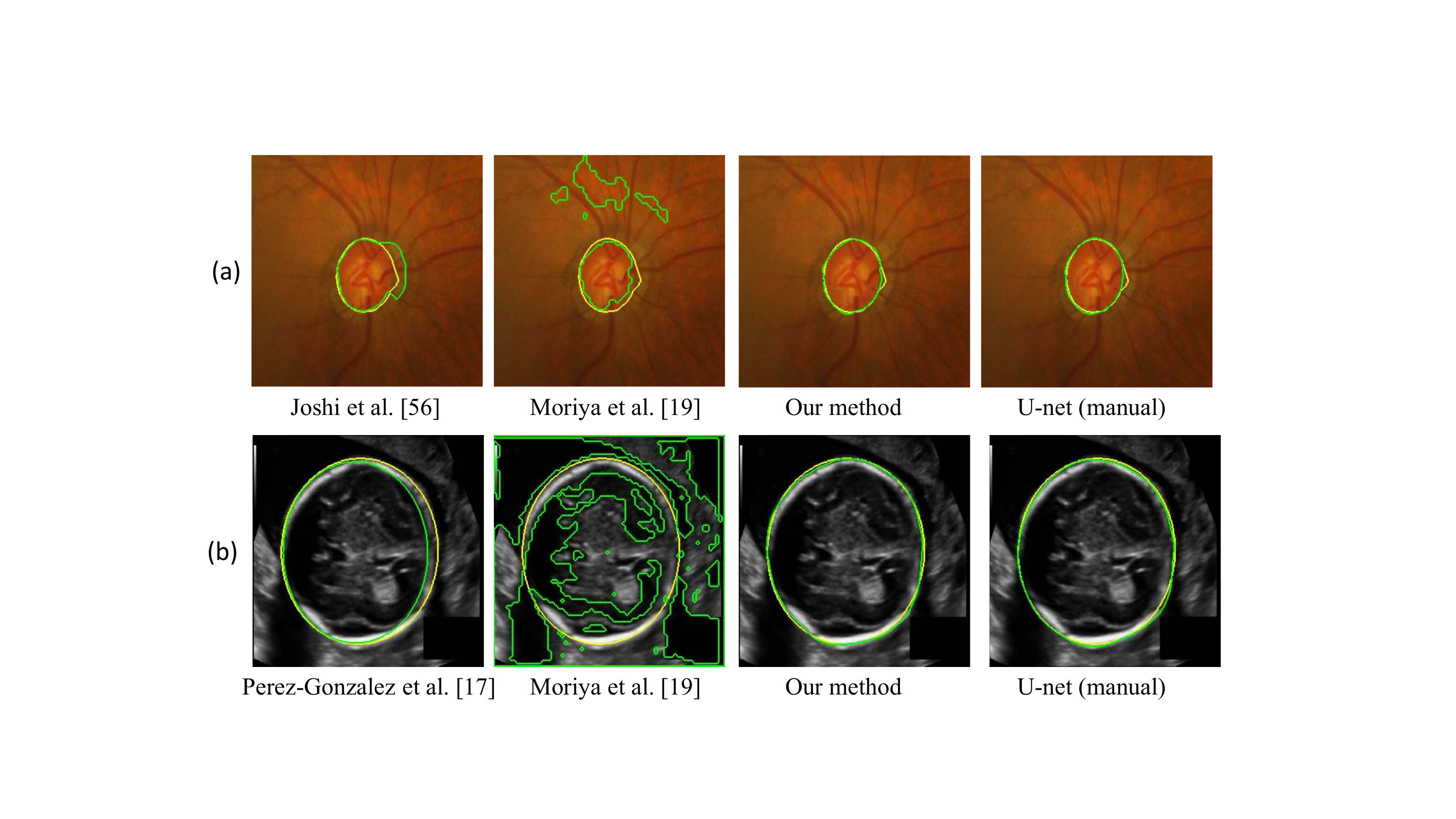}}
\caption{Visual comparison of our method with other methods. (a) and (b) show results for optic disc and fetal head, respectively. Green and yellow curves are segmentation results and the ground truth, respectively.}
\label{fig:fig9}
\end{figure}
\begin{figure}[!t]
\centerline{\includegraphics[width=1.00\columnwidth]{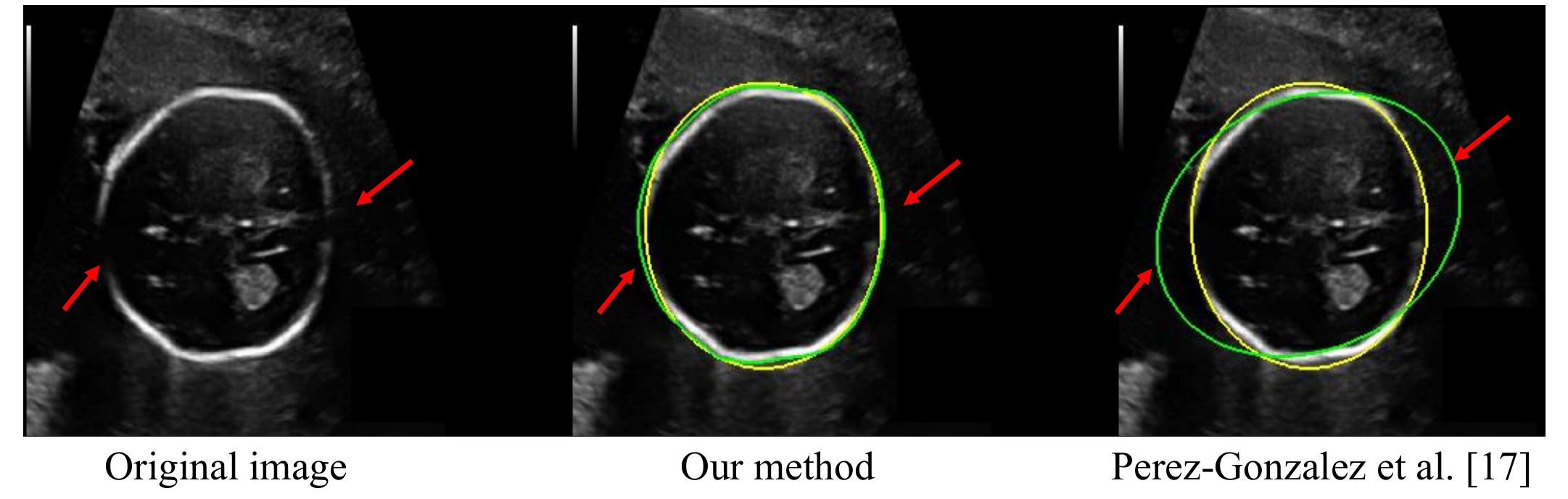}}
\caption{Visual comparison of our method with Perez-Gonzalez et al.~\cite{perez2015automatic} for fetal head segmentation when dealing with weak boundary information. Green and yellow curves are segmentation results and the ground truth, respectively. Red arrows highlight weak boundary information.}
\label{fig:figb}
\end{figure}
Our method was compared with U-Net (manual) that represents training U-Net with manual annotations, and it was compared with three existing unsupervised segmentation methods: 1) Joshi et al.~\cite{joshi2011optic} that uses circular Hough transform and snake model for optic disc segmentation; 2) The deep representation and adversarial learning-based method proposed by Moriya et al.~\cite{moriya2019unsupervised};  3) The method of Perez-Gonzalez et al.~\cite{perez2015automatic} that uses optimal ellipse detection and texture maps for fetal head segmentation. We also compared our method with two existing weakly-supervised methods: 1) Kervadec et al.~\cite{kervadec2019constrained} that employs a differentiable penalty in loss function to enforce inequality constraints; 2) Lu et al.~\cite{lu2019weakly} that trains a U-Net model with the foreground segmentation map generated by an improved constraint CNN  and GrabCut. Table~\ref{tab:noiseo} shows that our method largely outperformed existing unsupervised and weakly supervised methods for these two objects.

In the iterative training process, we also replaced our pseudo labels obtained by $G_A$ with those obtained by existing unsupervised methods, i.e., \cite{joshi2011optic} for optic disc and~\cite{perez2015automatic} for fetal head, which is denoted as~\cite{joshi2011optic} + IT + wDice and~\cite{perez2015automatic} + IT + wDice, respectively. Table~\ref{tab:noiseo} demonstrates that when pseudo labels obtained by~\cite{perez2015automatic} or ~\cite{joshi2011optic} are used, our iterative training still leads to a large performance improvement. However, the performance is worse than using pseudo labels obtained by our $G_A$ for the iterative training process. Table~\ref{tab:noiseo} also shows that the result of our method has no significant  difference from that of learning from human annotations for optic disc segmentation, and the performance gap is also subtle for fetal head segmentation. Visual comparison in Fig.~\ref{fig:fig9} shows that the result of our method is comparable to that of learning from human annotations and Fig.~\ref{fig:figb} shows that our method performs well when dealing with images with weak boundary information like fetal head segmentation.

\subsection{Segmentation of Complex Structures }
In this section, we apply our framework to complex structures where the shape prior can hardly be represented by a parametric model. To deal with this problem, instead of generating samples from a parametric shape model, we take advantages of a set of third-party segmentation masks that is available in public datasets. For the experiment, we consider the lung segmentation from Chest X-Ray (CXR) images and liver segmentation from CT images.

\subsubsection{Data}
For lung segmentation, we used the Japanese Society of Radiological Technology (JSRT) dataset~\cite{shiraishi2000development} that contains 247 posterior-anterior chest X-ray images with expert segmentation masks. The original images size was $2048 \times 2048$ with a pixel spacing of 0.715 mm $\times$ 0.715 mm. To learn without the annotations of JSRT images, we obtain auxiliary lung masks from the public Montgometry County X-Ray Set (MCXS)~\cite{jaeger2014two}. It contains 138 posterior-anterior CXR images, of which 80 images are normal and 58 images are abnormal with manifestations of tuberculosis. We used images in the JSRT dataset as domain $\mathbf{I}$ and lung masks in the MCXS dataset as domain $\mathbf{S}$ for our annotation-efficient learning.

For liver segmentation, we utilized the data form ISBI 2019 CHAOS Challenge~\cite{valindria2018multi}, which contains unpaired 20 CT volumes and 20 MRI volumes with expert segmentation masks. The CT volumes have an in-plane size of 512 $\times$ 512 and slice thickness of 1.5 mm. The MRI volumes have a size of 256 $\times$ 256 with varying slice thickness, and we re-sampled the slice thickness to 1.5 mm. Both CT and MRI images were cropped near the liver region in 3D, and slices without liver were excluded. We aimed to segment the liver from CT images (domain $\mathbf{I}$) with the help of auxiliary liver masks (domain $\mathbf{S}$) from the MRI images.

For these two applications, we randomly split the images of JSRT dataset and CT volumes of CHAOS dataset into 70$\%$ for training, 10$\%$ for validation and 20$\%$ for testing, respectively. And we abandoned the ground truth of the training images for our annotation-efficient learning. Each image was resized to $288 \times 288$, randomly cropped to 256 $\times$ 256, and the intensity was normalized into the range of [-1, 1].
\begin{figure}[!t]
\centerline{\includegraphics[width=1.00\columnwidth]{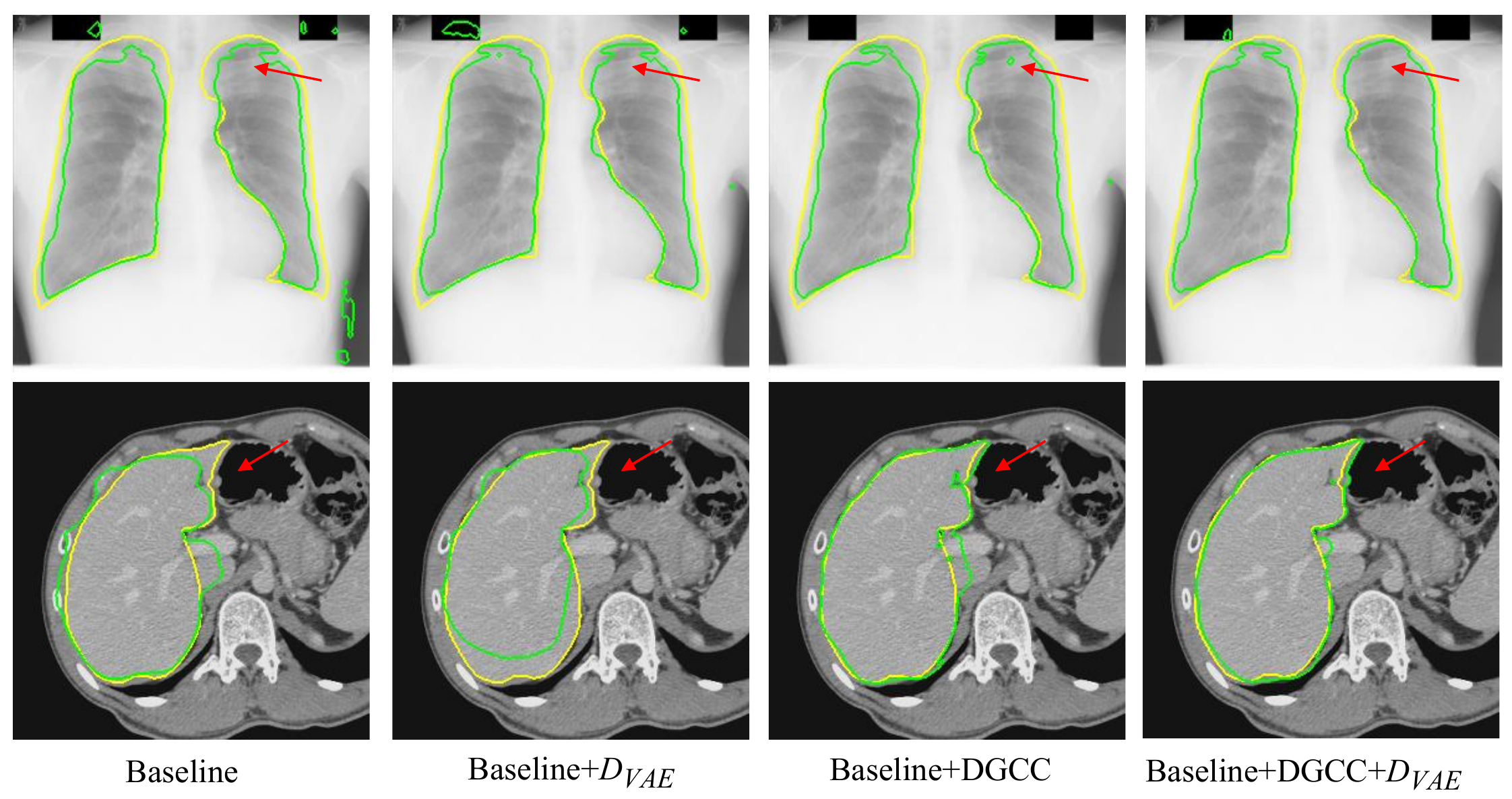}}
\caption{Visual comparison of results of pseudo label generator $G_A$ with different training methods for lung and liver. Green and yellow curves show the output of $G_A$ and the ground truth, respectively. Red arrows highlight some local difference.}
\label{fig:fig11}
\end{figure}
\begin{table}
	\centering
	\setlength{\tabcolsep}{3pt}
	\scriptsize
	\caption{Quantitative evaluation of the effect of $D_{V\!AE}$ and DGCC on the pseudo label generator $G_A$ for lung and liver segmentation.}
	\label{tab:lungp}
    \begin{tabular}{lcccc}
	\toprule
	\multirow{2}{*}{Methods} &\multicolumn{2}{c}{Lung} &\multicolumn{2}{c}{Liver} \\ \cline{2-5}
	&Dice & ASSD (pixel) &Dice & ASSD (pixel)\\ \hline
    Baseline & 0.823$\pm$0.056 & 11.681$\pm$3.798 & 0.853$\pm$0.099 & 8.023$\pm$5.630\\
    +$D_{V\!AE}$(w/o $D_{B}$)  & 0.799$\pm$0.051 & 12.953$\pm$2.975 & 0.837$\pm$0.096 & 10.122$\pm$4.546\\
    +$D_{V\!AE}$ & 0.839$\pm$0.045 & 9.955$\pm$3.345 & 0.869$\pm$0.044 & 7.451$\pm$3.806\\
    +DGCC & 0.833$\pm$0.047 & 10.436$\pm$3.536 & 0.865$\pm$0.093 & 6.862$\pm$4.783\\ 
    +$D_{V\!AE}$+DGCC & \textbf{0.848$\pm$0.046} & \textbf{9.406$\pm$3.592} & \textbf{0.889$\pm$0.062} & \textbf{5.656$\pm$1.901}\\ 
	\bottomrule
	\end{tabular}
\end{table}
\subsubsection{Effectiveness of VAE-based Discriminator and DGCC}
We first evaluate the effectiveness of our pseudo label generation method.
For ablation studies, we started with a baseline of training a CycleGAN~\cite{zhu2017unpaired} without $D_{V\!AE}$ and DGCC. It was compared with baseline+$D_{V\!AE}$, baseline+$D_{V\!AE}$ without $D_{B}$, baseline+DGCC, and baseline+$D_{V\!AE}$+DGCC.  Table~\ref{tab:lungp} lists quantitative evaluation results of the output of $G_A$ for our testing set. It shows that our $D_{V\!AE}$ improved the average Dice score from 0.823 to 0.839 for lung segmentation and 0.853 to 0.869 for liver segmentation compared with the baseline. Using $D_{V\!AE}$ and DGCC at the same time outperformed the other variants, with an average Dice score of 0.848 for lung segmentation and 0.889 for liver segmentation. Fig.~\ref{fig:fig11} shows a visual comparison between these variants. It can be observed that the output of $G_A$ trained by the baseline method contains some noise. By using $D_{V\!AE}$ that introduces a high-level shape constraint, the noise is reduced. DGCC also helps to improve the quality of $G_A$'s output compared with the baseline. The last column of Fig.~\ref{fig:fig11} shows that a combination of $D_{V\!AE}$ and DGCC obtained better results than the others.

\begin{figure}[!t]
\includegraphics[width=0.5\textwidth]{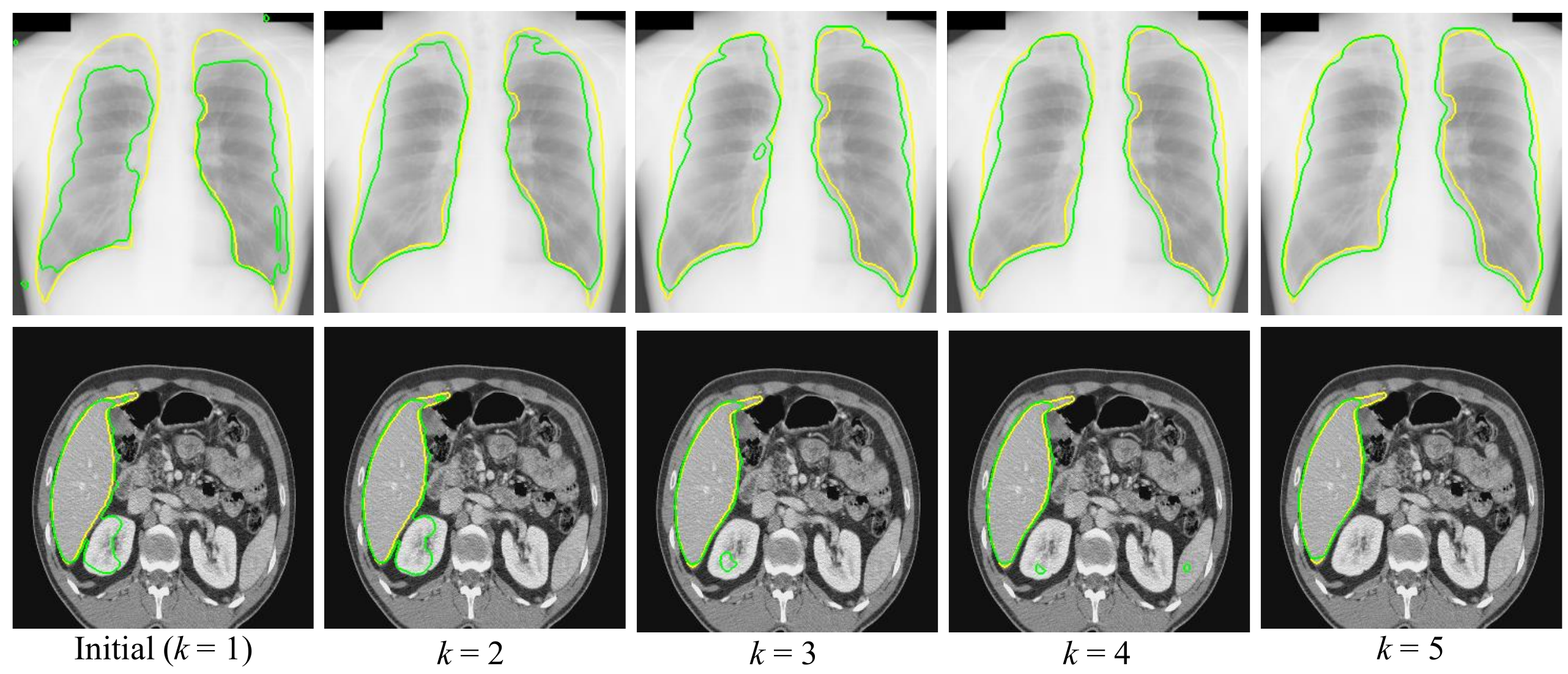}
\caption{Visual comparison of pseudo labels (green curves) for lung and liver segmentation at different rounds of our proposed iterative training method with noise-weighted Dice loss. Yellow curves are the ground truth label.}
\label{fig:fig12}
\end{figure}
\begin{figure*}[!t]
\centerline{\includegraphics[width=1.00\textwidth]{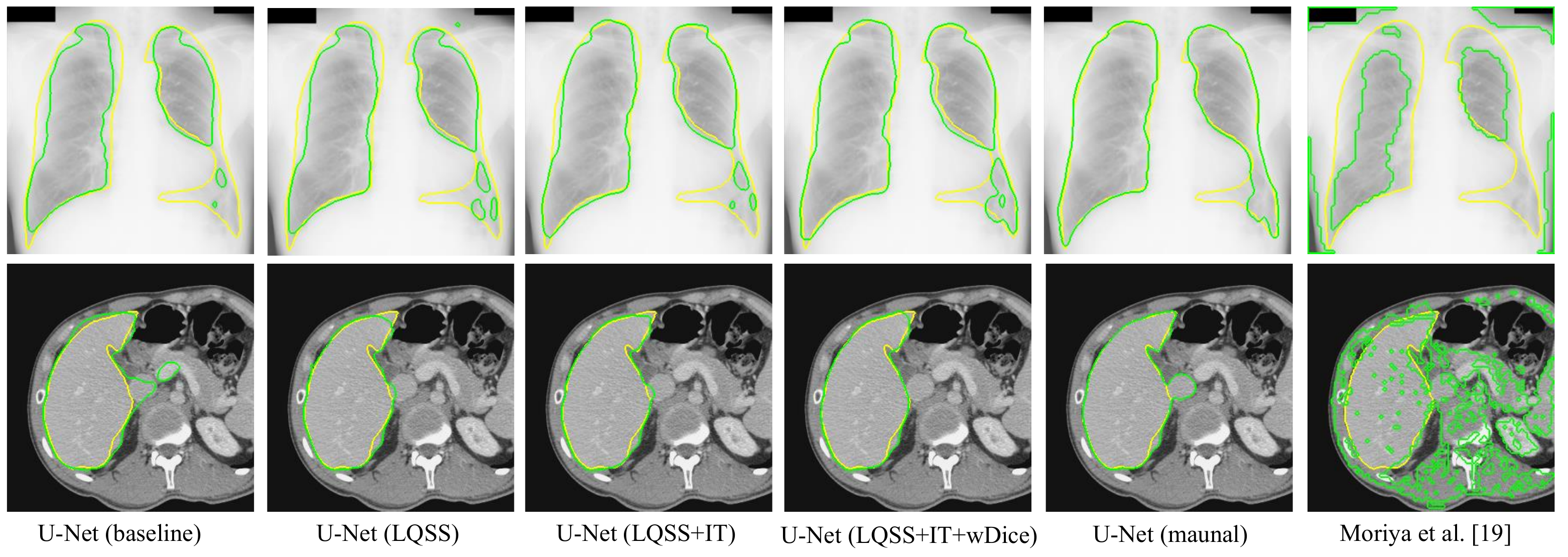}}
\caption{Visual comparison between different variants of our noise-robust method to learn from pseudo labels obtained by $G_A$ for lung and liver segmentation. U-Net (LQSS + IT + wDice) is our proposed method. U-Net (manual) is training from human annotations. Moriya et al.~\cite{moriya2019unsupervised} is an annotation-free method.}
\label{fig:fig13}
\end{figure*}
\begin{figure}[!t]
\centerline{\includegraphics[width=1.00\columnwidth]{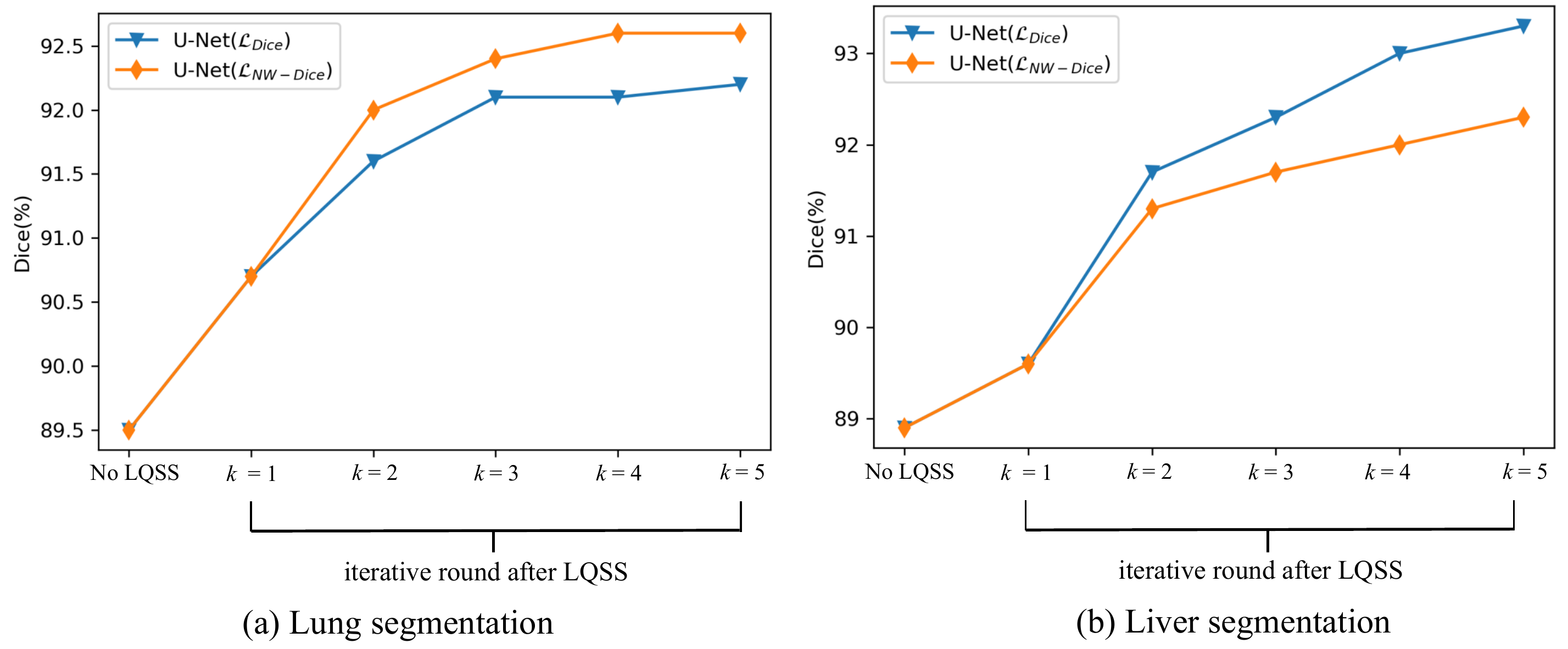}}
\caption{Effect of our proposed iterative training method for learning from pseudo labels obtained by $G_A$ for lung and liver segmentation. $k$ represents the number of rounds.}.
\label{fig:fig14}
\end{figure}
\begin{table}
	\centering
	\setlength{\tabcolsep}{1pt}
	\scriptsize
	\caption{Quantitative evaluation of different methods for lung and liver segmentation. Bold font highlights the best results obtained by our final segmentation model. $^\#$ denotes learning from pseudo labels obtained by our $G_A$. $^\square$ denotes learning from manual annotations. $^\triangle$ and $^\perp$ denote existing unsupervised and weakly supervised methods, respectively. $^\lozenge$ denotes unsupervised domain adaptation methods. }
	\label{tab:lungn}
    \begin{tabular}{lcccc}
	\toprule
	\multirow{2}{*}{Methods} &\multicolumn{2}{c}{Lung} &\multicolumn{2}{c}{Liver} \\ \cline{2-5}
	&Dice & ASSD (pixel) &Dice & ASSD (pixel)\\ \hline
    $^\#$U-Net (baseline) &0.895$\pm$0.040 & 5.462$\pm$2.322 & 0.896$\pm$0.064 & 4.896$\pm$1.869\\
	$^\#$U-Net (LQSS)  &0.907$\pm$0.031 & 4.687$\pm$1.607  &0.908$\pm$0.052 & 4.491$\pm$1.858\\
	$^\#$U-Net (LQSS+IT) & 0.922$\pm$0.029 & 3.997$\pm$1.396 &0.923$\pm$0.040 & 3.708$\pm$1.628\\ 
	$^\#$U-Net (LQSS+IT+wDice) &\textbf{0.926$\pm$0.025} & \textbf{3.693$\pm$1.178}  & \textbf{0.933$\pm$0.031} &\textbf{3.151$\pm$1.071}\\ \hline
	U-Net (no adapt.)&0.797$\pm$0.162 & 5.186$\pm$2.482 & 0.458$\pm$0.189 & 27.357$\pm$7.709\\
	$^\square$U-Net (manual)&0.947$\pm$0.024 &3.179$\pm$2.002 & 0.954$\pm$0.018 &2.414$\pm$0.869\\
    $^\lozenge$Wu et al.~\cite{9165963} &0.835$\pm$0.054 &6.826$\pm$2.169 &0.898$\pm$0.052 &4.607$\pm$1.985\\
	$^\perp$Kervadec et al.~\cite{kervadec2019constrained} & 0.820$\pm$0.064 & 8.620$\pm$5.056 &0.920$\pm$0.044 &4.013$\pm$1.726\\
	$^\triangle$Moriya et al.~\cite{moriya2019unsupervised} & 0.655$\pm$0.094 & 16.553$\pm$4.684  &0.516$\pm$0.121 &26.389$\pm$6.035\\
	\bottomrule
	\end{tabular}
\end{table}

\subsubsection{Results of Learning from Noisy Pseudo Labels}
To validate our noise-robust iterative method to learn from noisy pseudo labels obtained by our generator $G_A$, we first compared the following variants: 1) U-Net (baseline) that learns from the pseudo labels using a standard Dice loss without considering the existence of noise; 2) U-Net trained with Dice loss from samples selected by our LQSS, which is referred to as U-Net (LQSS). These two methods only train the model once without iterative training, and were further compared with: 3) U-Net (LQSS + IT) that refers to U-Net (LQSS) followed by iterative training with Dice loss (five rounds); and 4) U-Net (LQSS + IT + wDice) that refers to U-Net (LQSS) followed by iterative training with our noise-weighted Dice loss  (five rounds). The quantitative evaluation results are shown in Table~\ref{tab:lungn}. It can be observed that the LQSS obtained better performance than the baseline, and using iterative training and noise-weighted Dice loss can further improve the segmentation accuracy. As shown in Table~\ref{tab:lungn}, the iterative training with our noise-weighted Dice loss improved the segmentation Dice score from 0.907 to 0.926 for the lung, and from 0.908 to 0.933 for the liver, respectively. The results show that iteration process is important for learning from noisy pseudo labels. Fig.~\ref{fig:fig12} demonstrates the pseudo labels refined at different rounds of training stage, and it can be observed that the quality of pseudo labels are gradually improved during the training rounds. Fig.~\ref{fig:fig14} shows the performance of iterative training on the testing set. It demonstrates that the performance increased at the beginning and reached a plateau after four rounds, and also shows that noise-weighted Dice loss is better than Dice loss during the iterative training. 

\subsubsection{Comparison with Existing Methods and Learning from Human Annotations}
As our method uses auxiliary masks from a publicly available third-party dataset, we compared our method with applying the U-Net trained with the third-party dataset directly to our testing images, which is denoted as U-Net (no adapt.). As shown in Table~\ref{tab:lungn}, U-Net (no adapt.) has a poor performance for lung segmentation. This is because JSRT images are from normal persons while MCXS contains some abnormalities, and the two datasets have different intensity distributions, leading to a large domain shift. Similarly, U-Net (no adapt.) also performs poorly for the liver segmentation, due to the domain shift between MRI images and CT images. 

For comparison with training from full supervision, we trained U-Net with manual annotations of the JSRT dataset for lung segmentation and CT volumes of CHAOS dataset for liver semgentation, which is denoted as U-Net (manual). We also compared our method with an existing unsupervised method proposed by Moriya et al.~\cite{moriya2019unsupervised}, a weakly-supervised method proposed by Kervadec et al.~\cite{kervadec2019constrained} and an unsupervised domain adaptation proposed by Wu et al.~\cite{9165963} that uses a characteristic function distance metric based on characteristic functions of distributions to enable explicit domain adaptation. Table~\ref{tab:lungn} shows that our method largely outperformed that of Moriya et al.~\cite{moriya2019unsupervised}, Kervadec et al.~\cite{kervadec2019constrained} and Wu et al.~\cite{9165963}. What's more, both Table~\ref{tab:lungn} and Fig.~\ref{fig:fig13} show that the difference between our method and U-Net (manual) is subtle.

\section{DISCUSSION AND CONCLUSION}
\label{sec:guidelines}
A large set of high-quality manual annotations for medical image segmentation tasks is difficult and labor-intensive to acquire, which has been a crucial obstacle for developing deep learning methods. To alleviate this problem, some works have studied on annotation-efficient segmentation~\cite{bai2017semi,bortsova2019semi,toussaint2018weakly,feng2017discriminative,rajchl2016deepcut,kervadec2019constrained}, but they still require some annotations with human efforts for the training set. While there exist some previous works studying unsupervised (i.e., annotation-free) segmentation~\cite{moriya2018unsupervised,moriya2019unsupervised} through deep representation learning~\cite{radford2015unsupervised}, their accuracy for segmentation is limited. This paper proposes a new framework learning from a set of unpaired training images and auxiliary masks that can be easily obtained through either shape prior information or publicly available datasets in a probably different domain. With the help of auxiliary masks, we generate a pseudo segmentation label for each training image through our improved CycleGAN~\cite{zhu2017unpaired}, and pseudo labels are combined with our noise-robust learning process to get the final segmentation model.

Our improved CycleGAN can generate high-quality pseudo labels due to the following reasons. First, the auxiliary masks are based on either shape prior information or publicly available datasets, which provide a shape distribution of the target organ.  They are used by our adversarial networks to impose a shape constraint on the pseudo labels. Second, our DGCC module uses the feature map of the discriminator to directly calibrate the pseudo label generator for better performance. 

The advantage of our VAE-based discriminator is that it automatically learns a compact high-level shape representation, and it can be easily trained based on the auxiliary masks. The latent vector of VAE is an implicit modeling of the object shapes, which helps to constrain the generator. Despite the different shapes among organs we segment in this paper, the hyper-parameters of VAE were kept the same, i.e., vector length was 32 and $\lambda_{V\!AE}$ was 1.0. The results showed that such a setting is effective and general in all the four segmented organs. However, in other applications such as dealing with 3D images or segmenting vessels, these parameters may need to be tuned based on the specific dataset. One may replace the VAE by an encoder coupled with manual shape constraints. However, the latter relies on researchers’ experience, and effective manual constraints are hard to find for complex shapes.  

A segmentation model can be trained using pseudo labels obtained by our $G_A$. However, these labels are noisy and not very accurate. We overcome this problem by our noise-robust learning process, where pseudo labels and the final segmentation model are iteratively updated. By rejecting low-quality pseudo labels and weighting pixels according to the estimated noise level in Dice loss function for training, the effect of noisy labels is alleviated and thus a high-performance segmentation model can be obtained. Our noise-robust learning method may also be used for other situations where noisy labels exist, e.g., semi-supervised learning~\cite{bai2017semi} and learning from non-expert annotations~\cite{9109297}.

Our proposed framework can segment medical images without the expensive annotations for training images by taking advantage of the shape information from the auxiliary masks. Our basic assumption is that instead of annotations corresponding to training images, some auxiliary masks related to the target object class can be obtained without extra efforts. The auxiliary masks provide some shape information of the target and are not paired with the training set. We have shown that two possible ways to obtain such auxiliary masks: using a parametric shape model to generate a set of auxiliary masks for simple structures such as optic disc and fetal head, and taking advantage of masks of the object from another domain (e.g., public datasets) for complex structures such as the lung and the liver. For more complex structures such as the brain and vessels~\cite{wang2020augmenting}, it might be more challenging to leverage existing unpaired labels from a different dataset for shape constraint. The effectiveness of our method in such cases will be investigated in the future.  Our method in this paper is implemented by 2D networks, and theoretically it can also employ other  network structures and be extended to deal with 3D images.

In conclusion, we propose a novel annotation-efficient training framework for medical image segmentation by leveraging a set of auxiliary masks. An improved CycleGAN is proposed to learn from unpaired medical images and auxiliary masks, where adversarial learning leverages the auxiliary masks to introduce shape constraints on generated pseudo labels of training images. To improve the performance of pseudo label generator, we introduce a VAE-based discriminator and Discriminator-guided Generator Channel Calibration (DGCC). We also propose a noise-robust iterative training method to learn from the noisy pseudo labels, where a Label Quality-based Sample Selection (LQSS) module and a noise-weighted Dice loss are introduced to overcome noisy labels. Experimental results showed that our method achieved accurate segmentation results, which was close or even comparable to the same CNN structure trained with manual annotations. The framework provides a feasible solution for avoiding human annotations of training images, and we will investigate its application to segmentation of other structures in the future.


\bibliographystyle{IEEEtran}
\bibliography{IEEEreference}

\end{document}